\documentclass[12pt]{article}
\usepackage{booktabs}

\usepackage{newtxtext,newtxmath}

\usepackage{rotating}
\usepackage{caption}

\usepackage{graphicx}

\usepackage[letterpaper,margin=1in]{geometry}

\renewenvironment{abstract}
	{\quotation}
	{\endquotation}

\date{}

\makeatletter
\renewcommand{\fnum@figure}{\textbf{Figure \thefigure}}
\renewcommand{\fnum@table}{\textbf{Table \thetable}}
\makeatother

\usepackage{scicite}

\usepackage{url}

\def\scititle{
	Photo-generated charge-transfer excitons in NiO revealed by ultrafast time-resolved resonant inelastic x-ray scattering
}

\title{\bfseries \boldmath \scititle}

\author{
	Giacomo~Merzoni$^{1,2\ast\dagger}$,
	Leonardo~Martinelli$^{1\ast\dagger\ddagger}$,
	Sergii~Parchenko$^{2}$,\and
	Sophia~F.R.~TenHuisen$^{3}$,
        Vasily~Lebedev$^{4}$,
	Luigi~Adriano$^{2}$,
	Robert~Carley$^{2}$,\and
	Natalia~Gerasimova$^{2}$,
	Laurent~Mercadier$^{2}$,
	Martin~Teichmann$^{2}$,\and
	Benjamin~E.~van Kuiken$^{2}$,
	Zhong~Yin$^{2}$,
	Amina~Alic$^{5\S}$,
	Denitsa~R.~Baykusheva$^{3,6}$,\and
	Sorin~G.~Chiuzbaian$^{5}$,
	Stefano~Dal Conte$^{1}$,
	Oleg~Dogadov$^{1}$,
	Alexander~F\"ohlisch$^{7,8}$,\and
	Maurits~W.~Haverkort$^{9}$,
	Maximilian~Kusch$^{7}$,
	Tim~Laarmann$^{10,11}$,
	Wei~Sheng~Lee$^{12}$,\and
	Marco~Moretti Sala$^{1}$,
	Ying~Ying~Peng$^{13}$,
	Qing~Zheng~Qiu$^{13}$,\and
	Thorsten~Schmitt$^{14}$,
	Sreeju~Sreekantan Nair Lalithambika$^{10}$,
	Simone~Techert$^{10,15}$,\and
	Giulio~Cerullo$^{1}$,
	Michael~F\"orst$^{16}$,
	Matteo~Mitrano$^{3}$,
	Mark~P.M.~Dean$^{17}$,\and
	Justine~Schlappa$^{2}$,
	Andreas~Scherz$^{2\ast}$,
	Giacomo~Ghiringhelli$^{1,18\ast}$\and
	\small$^{1}$Dipartimento di Fisica, Politecnico di Milano, piazza Leonardo da Vinci 32, I-20133 Milano, Italy.\and
	\small$^{2}$European XFEL, Holzkoppel 4, D22869 Schenefeld, Germany.\and
	\small$^{3}$Department of Physics, Harvard University, Cambridge, Massachusetts 02138, USA.\and
        \small$^{4}$Bernal Institute, University of Limerick, Limerick, V94 T9PX, Ireland.\and
	\small$^{5}$Sorbonne Université, CNRS, Laboratoire de Chimie Physique - Matière et Rayonnement,\and \small  LCPMR, 75005 Paris, France.\and
	\small$^{6}$Institute of Science and Technology Austria, Klosterneuburg, Austria.\and
	\small$^{7}$Institute Methods and Instrumentation for Synchrotron Radiation Research, Helmholtz Center Berlin\and \small for Materials and Energy, 12489 Berlin, Germany.\and
	\small$^{8}$Institute of Physics and Astronomy, University of Potsdam, 14476 Potsdam, Germany.\and
	\small$^{9}$Institute for Theoretical Physics, Heidelberg University, Philosophenweg 19, 69120 Heidelberg, Germany.\and
	\small$^{10}$Deutsches Elektronen-Synchrotron DESY, 22607 Hamburg, Germany.\and
	\small$^{11}$The Hamburg Centre for Ultrafast Imaging CUI, 22607 Hamburg, Germany.\and
	\small$^{12}$Stanford Institute for Materials and Energy Sciences, SLAC National Accelerator Laboratory,\and \small2575 Sand Hill Road, Menlo Park, California 94025, USA.\and
	\small$^{13}$International Center for Quantum Materials, School of Physics, Peking University, Beijing 100871,\and \small People's Republic of China.\and
	\small$^{14}$PSI Center for Photon Science, Paul Scherrer Institut, 5232 Villigen PSI, Switzerland.\and
	\small$^{15}$Institut für Röntgenphysik, Georg-August-Universität Göttingen, 37077 Göttingen, Germany.\and
	\small$^{16}$Max Planck Institute for the Structure and Dynamics of Matter, 22761 Hamburg, Germany.\and
	\small$^{17}$Condensed Matter Physics and Materials Science Division, Brookhaven National Laboratory, Upton,\and \small New York 11973, USA.\and
	\small$^{18}$CNR-SPIN, Dipartimento di Fisica, Politecnico di Milano, piazza Leonardo da Vinci,\and \small I-20133 Milano, Italy.\\
    \small$^\ast$Corresponding authors. Emails: giacomo.merzoni@polimi.it, leonardo.martinelli@physik.uzh.ch,\\ \small andreas.scherz@xfel.eu, giacomo.ghiringhelli@polimi.it\and
	\small$^\dagger$These authors contributed equally to this work.\and
    \small$^\ddagger$ Current address: Physik-Institut, Universität Zürich, Winterthurerstrasse 190, CH-8057 Zürich, Switzerland.\and
    \small$^\S$ Current address: Université Sorbonne Paris Nord and Université Paris Cité,\and \small INSERM, LVTS, F-75018 Paris, France.
}

\begin{document} 

\maketitle

\begin{abstract} \bfseries \boldmath
Strong electronic correlation can lead to insulating behavior and to the opening of large optical gaps, even in materials with partly filled valence shells. Although the non-equilibrium optical response encodes both local (quasi atomic) and collective (long range) responses, optical spectroscopy is usually more sensitive to the latter. Resonant x-ray techniques are better suited to investigate the quasi-atomic properties of correlated solids. Using time-resolved resonant inelastic x-ray scattering (RIXS), here we study the ultrafast non-equilibrium processes in NiO following photo-excitation by ultraviolet photons  with energy exceeding the optical gap. We observe the creation of charge-transfer excitons that decay with a time constant of about 2\,ps, while itinerant photo-doping persists for tens of picoseconds. Following our discovery, which establishes time-resolved high-resolution RIXS as a powerful tool for the study of transient phenomena in condensed matter, the possible presence of charge-transfer excitons will need to be considered when interpreting optical pump-probe experiments on correlated quantum materials. 
\end{abstract}

\noindent

\section{Introduction}\label{sec1}

\begin{figure}
\centering
\includegraphics[width=1\textwidth]{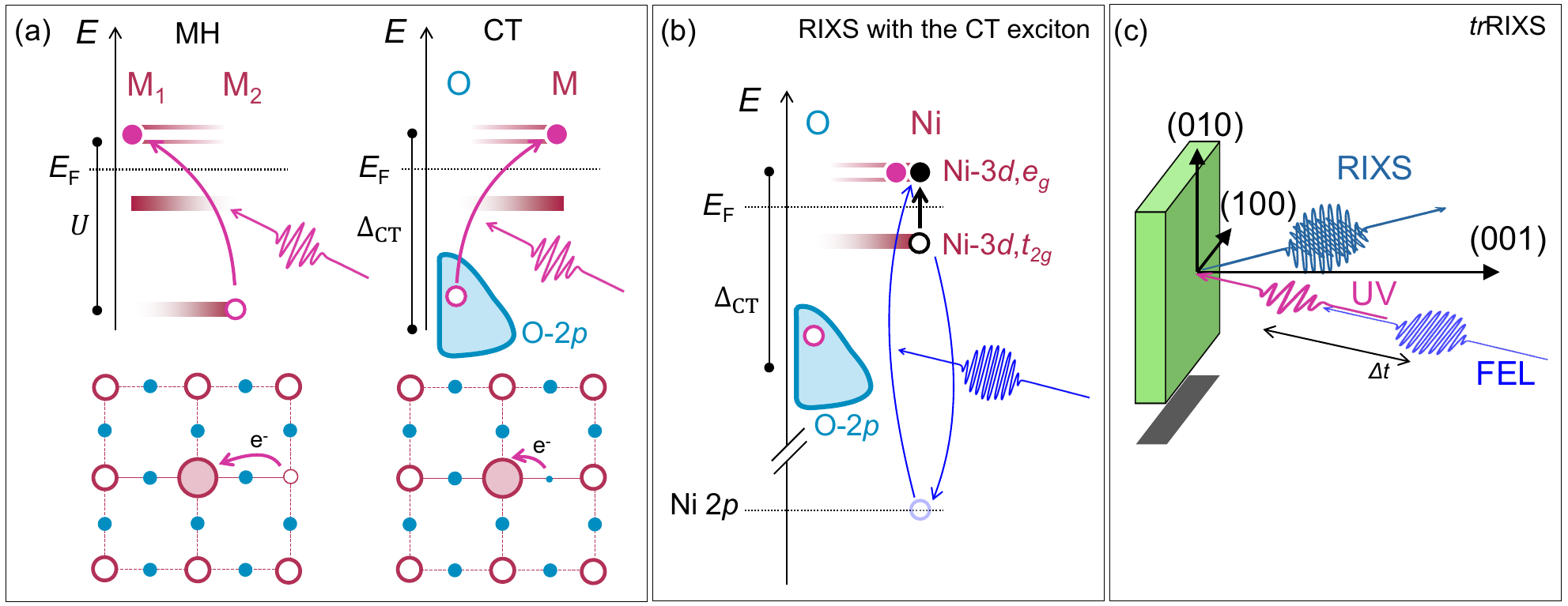}
\caption{\label{fig0}(a) Cartoon of the photo-excitation process across the correlation gap in Mott-Hubbard (MH) and charge-transfer (CT) insulators, in the energy (top) and the real space (bottom). M$_i$ indicate metal sites (e.g. a $3d$ transition-metal), O  the oxygen ligand, $U$ the on-site Coulomb repulsion and $\Delta_{\mathrm{CT}}$ the charge-transfer energy. (b) For the specific case of NiO, the same scheme is used for the RIXS process in the presence of a CT exciton. The ligand hole and photo-excited electron are represented by the empty and filled circles, respectively, as in panel (a). The resonant character of RIXS enhances the relation of the measurement with the local Ni $3d$ states. The black arrow indicates generic $dd$ excitations resulting from the RIXS process. (c) Experimental geometry of our \textit{tr}RIXS: a UV pulse photo-excites the sample and a collinear x-ray pulse impinges on it with a controllable delay. Inelastically scattered x-rays are analyzed by a grating spectrometer (not shown).}
\end{figure}

The optical constants of solids encode a wealth of information about the underlying electronic and magnetic structure. When the electron-electron interaction cannot be neglected, modeling optical constants becomes challenging though extremely valuable. In this context, experiments investigating the transient effects of light pulses on these materials are crucial for establishing connections to microscopic physics. They also create opportunities to drive the system into states that are unreachable in static conditions. That explains why ultrafast pump-probe techniques have become a primary tool for the investigation of quantum materials, enabling the realization of transient and eventually meta-stable states.
Remarkable optically-driven phenomena have been observed, such as transient phase transitions \cite{cavalleri2001femtosecond,polli2007Coherent}, the quenching of ordered phases \cite{boschini2018Collapse,Nevola_meleted_Fe_based,mitrano2019ultrafast,dean2016Ultrafast} and the photo-induced renormalization of characteristic effective interaction parameters, such as the Coulomb repulsion $U$ of the Hubbard model, the magnetic inter-atomic superexchange $J$, and the charge-transfer energy $\Delta$ \cite{Mitrano_XAS_U_renorm,Eschenlohr_NiO_CT_XAS,Granas_NiO}.
The lowest energy dipole-allowed transition is a metal-to-metal charge-transfer in Mott-Hubbard insulators, while a ligand-to-metal electron transfer in charge-transfer (CT) insulators, as sketched in Figure~\ref{fig0}(a) \cite{ZSA}. The former was predicted theoretically \cite{Clarke_MHExc,Lenarcic_MHExc,Huang_MHE,Shinjo_MHE,Wrobel_MHE} and proved experimentally \cite{mehio2023hubbard} to drive the formation of holon-doublon pairs in the form of so-called Hubbard excitons; instead, it is unclear whether the latter leads to bound electron-hole pairs (CT excitons) or itinerant states or both.
Despite their relevance to widely studied materials, e.g. high-Tc superconducting cuprates \cite{Hanamura_CTE}, such mechanisms have not been disclosed yet.
Our aim is to determine experimentally whether the CT exciton, corresponding to the Hubbard exciton, exists for CT insulators, how fast it decays and, possibly, through which processes.

On the one hand, state-of-the-art all-optical pump-probe techniques rely on the measurement of optical constants in the photo-excited states, therefore weighting more the collective, non-local part of the electronic structure with respect to the local, quasi atomic-one. On the other hand, resonant x-ray spectroscopy is known to be well suited to study the electronic properties from a local point of view \cite{Ament_review,Mitrano_review}. The combination of time-resolved methods with resonant x-ray spectroscopy can therefore provide a very original vision of the transient physics of quantum materials \cite{Mitrano_review}. Here we exploit this unique capability by using time-resolved resonant inelastic x-ray scattering (\textit{tr}RIXS) with high energy resolution, described in Figure~\ref{fig0}(b), to investigate the existence and the dynamics of CT excitons. In fact, static RIXS has been extensively used to obtain significant results on orbital \cite{Ghiringhelli_2005,Moretti_Sala_2011,Schlappa_orbitons_2012,hepting2020Electronica}, spin \cite{braicovich2010magnetic,LeTacon_paramag_2011,PengInfluence_2017,WSL_ILN_magnons_2021}, lattice \cite{Rossi_phonons_2019,Vale_iridate_phonons_2019} and charge excitations \cite{chaix2017dispersive,lee2021spectroscopic,ArpaiaScience_2019,HuangCDF_2021} of $3d$ transition-metal oxides. The availability of high repetition rate x-ray free electron  lasers now allows pushing a photon-hungry technique like RIXS to femtosecond temporal resolution.
We selected NiO, a reference material with a large correlation gap, simple crystalline structure and a well-known antiferromagnetic order, because it has been extensively studied with resonant x-ray spectroscopy  \cite{Wrobel_Mottness_NiO,newman1959optical,finazzi_resAES,Tjernberg_resonant_PES_NiO,Sawatzky_DFT_NiO}, and RIXS in particular \cite{Ghiringhelli_2005, ghiringhelli2009observation, betto2017three, nag2020many}.
In the face-centered cubic crystal structure of NiO, the Ni$^{2+}$ ions are octahedrally coordinated with O$^{2-}$ ligands and host ($S=1$, $L=0$) atomic moments that, below $T_{\mathrm{N}}=532$\,K, order in a collinear antiferromagnetic lattice. The strong Coulomb repulsion entails that the transfer of an electron from a metal site to a neighboring one costs several eV in energy ($U>6$\,eV) \cite{Hufner_NiO}. The Ni$3d^8 4sp^0$-O$2p^6$ ground state is thus well described by a $O_h$ ligand-field model, which is dominated by atomic multiplet states of Ni $3d^8$, with $^3$A$_{2g}$ symmetry and $(t_{2g}^6 \, e_g^2)$ orbital occupation. The resulting CT insulator has an optical gap about 4\,eV wide  \cite{Ghiringhelli_2005}.

Most pump-probe measurements on NiO reported so far have  used in-gap optical excitation \cite{Granas_NiO,Tzschaschel_PhysRevB.95.174407,Takahara_PhysRevB.86.094301,Satoh_PhysRevB.74.012404,Satoh_PhysRevLett.105.077402,Duong_PhysRevLett.93.117402,Bossini_PhysRevLett.127.077202}, whereas excitations above the CT gap and their evolution on the sub-picosecond timescale have been rarely investigated \cite{Eschenlohr_NiO_CT_XAS,gillmeister2020ultrafast}, possibly also due to the complexity of describing theoretically the many-body state in a photo-excited correlated system.
To study the transient properties of these excited states, we illuminate the $(001)$ face of a NiO single crystal with laser pulses of photon energy \mbox{$h\nu=4.66$ eV} ($\lambda=266$\,nm) exceeding the optical gap, and probe it by \textit{tr}RIXS. Since the photo-excitation changes the valence of the ion hosting the exciton, we exploit the unique chemical sensitivity of RIXS to isolate the details of the electronic properties of the photo-excited sites \cite{Kunnus_2016} and to follow their dynamics, as well as to monitor how the Ni sites not hosting the CT exciton are perturbed both by the presence of CT excitons nearby and by itinerant charges.

\section{Results}\label{sec2}
\begin{figure}
\centering
\includegraphics[width=1\textwidth]{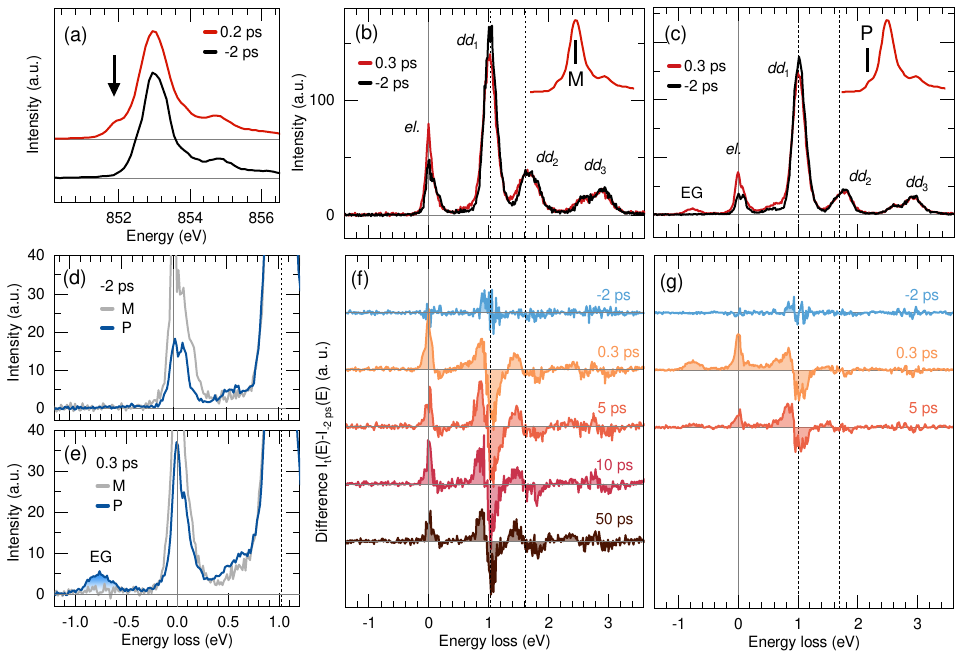}
\caption{\label{fig:exp_results} Time-resolved Ni L$_3$ XAS and RIXS spectra of NiO. (a) \textit{tr}XAS of NiO, measured at \mbox{0.2 ps} delay (red) and compared to that at negative time-delay (-2\,ps, black). (b,c) \textit{tr}RIXS spectra at two incident energies, on the main Ni$^{2+}$ resonance (853 eV, M) and on the transient XAS pre-edge (851.8 eV, P) as indicated in the insets by the vertical line on the XAS at 0.2 ps. In red we plot the RIXS spectra at 0.3\,ps delay, i.e., after the photo-excitation pulse; in black the reference spectra at negative delay (-2\,ps). (d) Comparison of the unperturbed (negative delay) RIXS spectra at the two incident energies, in the low energy scale (black spectra in panels (b,c). (e) Same comparison of panel (d) for 0.3 ps delay (red spectra in panels (b,c)). The energy gain feature is labeled EG. (f,g) Difference RIXS spectra at selected delays measured with M and P incident x-ray energy, respectively. The EG peak at -0.75\,eV appears only at 0.3\,ps. The main $dd$-excitation peak at -1\,eV (\textit{dd}1 in panels b and c), is shifted to lower energy-loss for all positive delays, resulting in derivative-like shape in the difference curves of panels (f,g).}
\end{figure}

We use time-resolved x-ray absorption spectroscopy (\textit{tr}XAS) and \textit{tr}RIXS, taking advantage of the well-established data interpretation framework available for the specific case of NiO \cite{Ghiringhelli_2005,nag2020many,quanty_paper}. \textit{tr}XAS and \textit{tr}RIXS spectra are acquired at the Ni $L_3$ edge, corresponding to the resonant core excitation $2p \rightarrow 3d$ or, in terms of electronic configurations, $2p^63d^n\rightarrow2p^53d^{n+1}$, where $n=8$ in the ground state but can be different in the photo-excited state. 
The static and photo-excited XAS spectra of  NiO are shown in Figure~\ref{fig:exp_results}(a). The static spectra are composed of a main resonance at $853$\,eV and a satellite at $854.5$\,eV; they are the two dominant terms of the $2p^53d^9$ multiplet, having spin-triplet and spin-singlet character, respectively \cite{Ghiringhelli_2005}. A pre-edge feature at $\sim$851.8\,eV, i.e. $\sim\!1.2$ eV below the main peak, appears in the XAS measured 0.2\,ps after the pump-pulse, in agreement with  reference~\cite{Eschenlohr_NiO_CT_XAS} but more evident due to the higher pump-pulse fluence used here ($\sim9.1$ mJ/cm$^2$, see supplementary materials). By referring to the static XAS of infinite-layer nickelates like NdNiO$_2$, where Ni is nominally mono-valent ($3d^9$) \cite{hepting2020Electronica}, and assuming a hole at the oxygen ligand after the photo-excitation, we can assign the pre-edge peak to the absorption of x-rays by a Ni site being momentarily in the  $3d^9$\underline{L} configuration (where \underline{L} stands for a hole in the ligand $2p$ states), instead of the ground state $3d^8$. We show below that calculations confirm this assignment. Figures~\ref{fig:exp_results}(b,c) show the \textit{tr}RIXS spectra excited at the main resonance (M excitation) and pre-edge peak (P excitation), respectively. The RIXS spectra are dominated by \textit{dd} excitations (reshuffling of the electrons among the $3d$ orbitals, also named crystal-field or ligand-field excitations) in the 1\,eV to 3\,eV energy-loss range \cite{Ghiringhelli_2005}, and by magnons and double-magnons ($\Delta S=1,2$ respectively) in the 0.1\,eV to 0.4\,eV interval \cite{ghiringhelli2009observation, betto2017three, nag2020many}. The dominant peak at $\sim$1\,eV, labeled $dd_1$, is assigned to the $^3$T$_{2g} \: (t_{2g}^5 \, e_g^3)$ final state and its energy  equals the value of the octahedral ligand-field parameter $10Dq\simeq\!1$\,eV~\cite{newman1959optical}. Additional peaks at higher energy-loss at 1.8\,eV, 2.5\,eV and 3.0\,eV  correspond to multiplet terms with $^3$T$_{1g}$, $^1$T$_{2g}$, $^1$A$_{1g}$ symmetry, respectively \cite{Ghiringhelli_2005}. 

When compared to the reference RIXS spectrum measured before the arrival of the pump-pulse at -2\,ps delay, and thus akin to the static one, the \textit{tr}RIXS spectra at positive delays and excited at the main absorption resonance display changes of all features up to 50\,ps. The quasi-elastic intensity is enhanced and the $dd$ peaks soften, as revealed by the derivative-like shape of the difference spectra in Figure~\ref{fig:exp_results}(f). Spectra excited at the absorption pre-edge feature are quite different from those of the main XAS peak: two extra features appear at +0.6\,eV and -0.75\,eV for short delays. The latter corresponds to an energy gain (EG) in the x-ray scattering process. The comparison of the RIXS spectra measured at the two incident energies (M for the main peak and P for the pre-edge feature in the XAS) is presented in Figures~\ref{fig:exp_results}(d,e) for the -2\,ps and 0.3\,ps delay, respectively.
Although all spectral modifications appear instantaneously within the experimental time resolution of about 100\,fs (see Methods for details), it is easy to see in Figures~\ref{fig:exp_results}(f,g) that the recovery time constants exceed $50$\,ps for the $dd_1$ and the quasi-elastic peaks, while they are shorter than 5\,ps for the EG peak. We must highlight here that via the choice of the incident energy we are selecting either the photo-excited Ni sites (P excitation) or the unperturbed ones (M excitation). Thus the \textit{tr}RIXS spectra show that the above-gap photo-excitation branches into two coexisting intermediate processes: a fast-decaying transient state, whose atomic multiplet is different from that of the ground state and whose signature is represented by the EG feature, and a long-lived perturbation of the crystal-field, monitored by the softening of the $dd$ excitations. 

\section{Discussion}\label{sec3}
\begin{figure}
\centering
\includegraphics[width=1\textwidth]{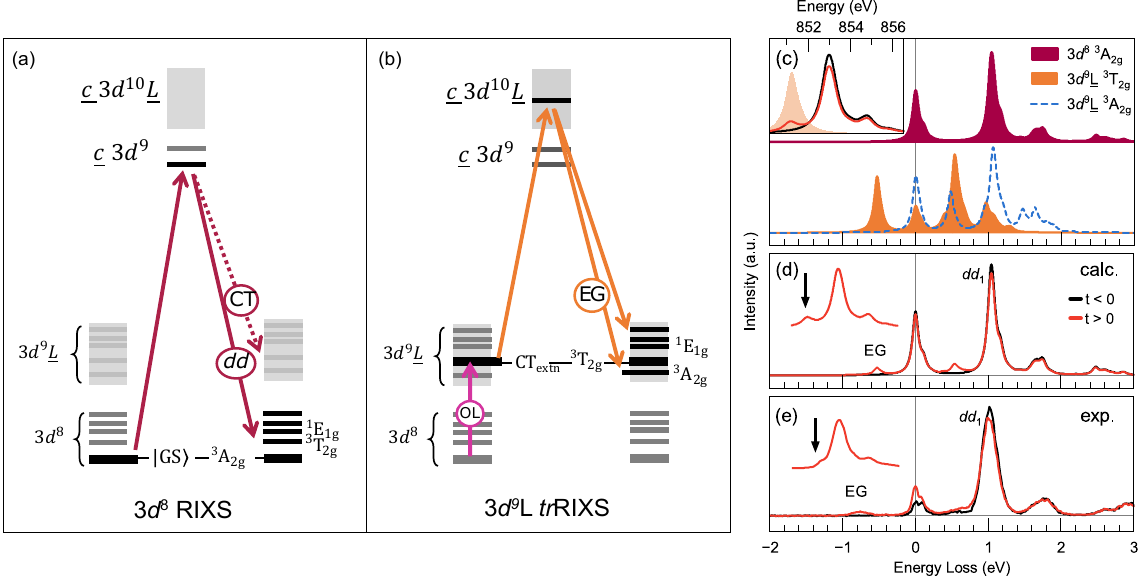}
\caption{\label{fig:calculations} Theory of \textit{tr}RIXS of NiO in the single-ion model with ligand-field. (a) Total energy scheme for the 3$d^8$ configuration, appropriate for static RIXS and for \textit{tr}RIXS with M excitation. The two largest cross sections are for $dd$ excitations that correspond to transitions from the $^3$A$_{2g}$ ground state to $^3$T$_{2g}$ and $^1$E$_{1g}$ states, at about 1.0\,eV and 1.8\,eV above it, respectively. (b) \textit{tr}RIXS with P excitation, which selects sites hosting a CT exciton: the initial state of both XAS and RIXS is the $^3$T$_{2g}$ term of the $3d^9$\underline{L} multiplet, populated by the optical laser (OL) photo-excitation. One of the dominant final states, $^3$A$_{2g}$, is lower in energy than the initial state, leading to the EG feature.  (c) Calculated RIXS and XAS (inset) spectra. The total XAS (red line) is obtained as linear combination of the standard one (3$d^8$ 
 ground state, black line) and the XAS of a Ni site hosting a CT exciton (3$d^9$\underline{L} configuration, orange shaded curve), as explained in the methods section. The standard RIXS ($3d^8$ multiplet) is the   magenta shaded curve at the top, the spectra for a site hosting the CT exciton are the orange shaded ($^3$A$_{2g}$) and the blue dashed ($^3$T$_{2g}$) line at the bottom. (d) The total RIXS spectrum excited at the pre-edge peak (inset) in the presence of CT excitons (red solid line) is obtained as the weighted average of the magenta and orange curves of panel c. (e) Experimental \textit{tr}XAS (inset) and \textit{tr}RIXS with P excitation for -2\,ps (black line) and +0.3\,ps (red line) delay.}
\end{figure}

We start our analysis from the main effects, the pre-edge peak of \textit{tr}XAS and the EG feature of \textit{tr}RIXS. A valuable insight is provided by ligand-field single-ion calculations that can reproduce very well the static XAS and RIXS spectra of NiO \cite{quanty_paper}. In that model the CT photo-excited state is described by a localized Ni 3$d^9$\underline{L} configuration.
Figure~\ref{fig:calculations}(a) shows a sketch of the static RIXS process, with the initial and final states all belonging to the $3d^8$ multiplet. The final states lie at an energy higher than the initial state (ground state) and correspond to the set of $dd$ excitations. Similarly, for \textit{tr}RIXS with P excitation, we assume that the initial and final states belong to the 3$d^9$\underline{L} configuration, as shown in Figure~\ref{fig:calculations}(b). This scheme is valid in the assumption that the Ni scattering site is hosting a localized CT exciton. 

The simulated \textit{tr}XAS shown in the inset of Figure~\ref{fig:calculations}(c) (red line) is the weighted sum of a conventional XAS calculated for the $3d^8,^3$A$_{2g}$ ground state and of a XAS process for a $3d^{9}$\underline{L}$,^3$T$_{2g}$ initial state. The excitonic pre-edge peak observed experimentally is reproduced remarkably well by weighting the two theoretical spectra according to the estimated density of excited sites in our experimental conditions ($\sim 10\%$ of the Ni sites host an exciton at very short delays, see Methods for details). 

The RIXS simulations shown in Figure~\ref{fig:calculations}(c,d) follow the same approach. It is worth noting that the $3d^{9}$\underline{L} configuration shares the same multiplet structure as the $3d^{8}$ ground state: 2 holes in total, one on Ni and one on the ligand as for $3d^{9}$\underline{L}, both on Ni as for $3d^{8}$, but with the same symmetry of the many-body solutions. The case of infinite-layer nickelates with pure $3d^{9}$ configuration is different, because there is only one hole in the configuration.
The calculated \textit{tr}RIXS spectrum in Figure~\ref{fig:calculations}(d) shows all the main features of the experimental \textit{tr}RIXS with P excitation. Its overall shape is similar to the static RIXS but with the addition of the EG peak and of the feature at $\sim0.6$\,eV. Those are the signatures of the $3d^9$\underline{L} configuration, i.e., of the localized CT exciton created by the absorption of the UV pump photon. It must be highlighted that the EG peak is necessarily due to the choice of the $^3$T$_{2g}$ term, which is not the lowest energy term in the multiplet, as initial state of the RIXS process. Indeed, the RIXS calculated for the $^3$A$_{2g}$ initial state, which is the lowest energy term of the $3d^9$\underline{L} manifold, does not have EG peaks, as shown by the blue dashed line at the bottom of Figure~\ref{fig:calculations}(c).  We note that the two $3d^9$\underline{L} spectra calculated for the $^3$T$_{2g}$ and $^3$A$_{2g}$ initial states cannot be superimposed by a rigid shift in energy, because they are not identical in shape. In Figure~\ref{fig:calculations}(e) the experimental \textit{tr}RIXS at the pre-edge is shown for comparison. Although the energy of the calculated EG peak does not match exactly the experiment, the agreement is remarkably good when considering that we used for the $3d^9$\underline{L} case the very same set of parameters previously optimized for the static RIXS of the $3d^8$ configuration. The discrepancy might be due to non-local screening effects and renormalization of characteristic energy scales in the photo-excited state that can be easily reproduced by our localized model  with an ad hoc tuning of the parameters, as shown in Figure~\ref{Afig:opt_calc} of the supplementary materials.
Finally, we highlight that the peak at $\sim0.6$\,eV in the calculated spectra corresponds well to the extra feature that appears around that energy in the \textit{tr}RIXS with P excitation: it is another term of the 3$d^9$\underline{L} multiplet, indicated by the dashed line in Figure~\ref{fig:calculations}(c-bottom).

\begin{figure}
\centering
\includegraphics[width=0.7\textwidth]{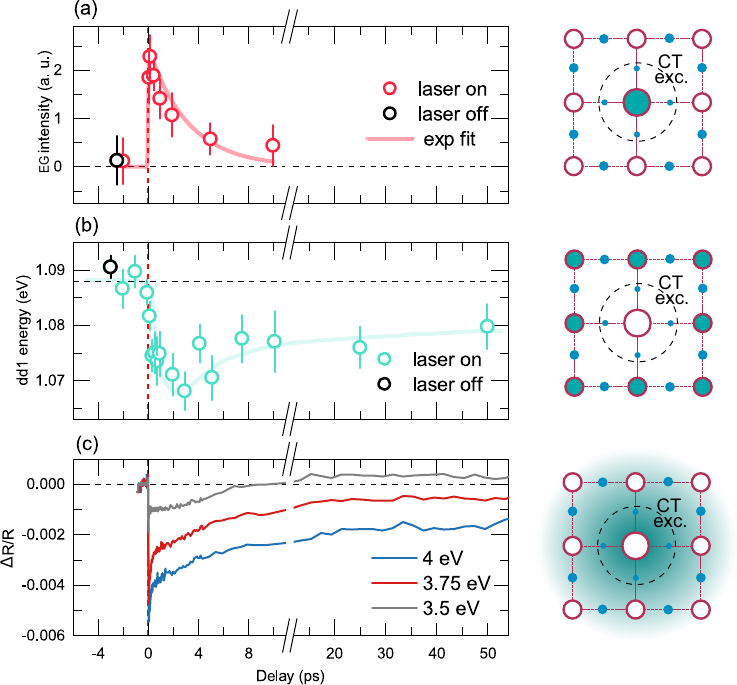}
\caption{\label{fig:dynamics} Time traces from \textit{tr}RIXS and optical transient reflectivity of NiO.  (a) Integrated intensity of the EG peak versus time delay detected in spectra measured with P excitation energy. The gray line represents a single exponential fitting convoluted by the experimental time resolution: the time constant is $\sim2$ ps. In the cartoon the green circle marks a Ni site scattering x-rays, the dashed circle the location of the CT exciton. (b) The $dd1$ peak energy as function of the delay. The faint line is a guide to the eye. In the cartoon the green colored  Ni ions  act as scattering sites but do not host the CT exciton. Itinerant charge and/or nearby excitons lead to the $dd1$ peak softening. (c) Transient vis-UV optical reflectivity with 4.66 eV laser pump- and 3 probe-photon energies (4.0\,eV, 3.75\,eV, 3.5\,eV). Following the initial fast response, the recovery does not follow a single exponential decay and different time constants of few ps and hundreds of ps are apparent. The cartoon shows that the optical reflectivity mostly probes non-localized states and that $\Delta R / R$ is mainly sensitive to modifications of the itinerant charge density.}
\end{figure}

Having interpreted the \textit{tr}RIXS spectra in the first 2\,ps using a local model, we can track the dynamics of the excitons by looking at the time dependence of the EG peak intensity (Figure~\ref{fig:dynamics}(a)) and of the $dd_1$ peak softening (Figure~\ref{fig:dynamics}(b)). Indeed, while we can argue that  photo-excitation above the optical gap in NiO is initially dominated by the creation of CT excitons, we also observe that the ligand-field is re-normalized by $\sim20$\,meV at all positive delays up to 50\,ps. To understand this result, we propose two distinct scenarios. The softening and broadening of \textit{dd} excitations is often observed upon hole-doping in many correlated oxides, such as cuprates and nickelates \cite{Huang_dd_screening,merzoni2023charge,fumagalli2019polarization,braicovich2010magnetic}. It is attributed to the screening action of itinerant positive carriers added to the ligand states. We can thus consider the \textit{dd} softening as the signature of a photo-induced delocalized hole-doped state in a correlated insulator, necessarily accompanied by negative charge promoted to the conduction band. As a different scenario, we notice that the increase of the lattice temperature often causes a softening and a broadening of the \textit{dd} excitations \cite{merzoni2023charge, barantaniPRXdd}. In our \textit{tr}RIXS experiment, the rapid temperature increase might result from the energy transfer from the electron bath to the lattice via electron-phonon scattering. Although distinguishing between the two phenomena at long delays is difficult, we highlight that the almost instantaneous rise time of the initial response of the \textit{dd} excitations to the optical pump seems too fast to be attributed only to thermalization in an insulator \cite{giannetti2016review}.

An insight comes from all-optical pump-probe spectroscopy. In Figure~\ref{fig:dynamics}(c) we present selected ultrafast differential reflectivity ($\Delta R / R$) measurements in the visible and ultraviolet (vis-UV). The full vis-UV map obtained with the supercontinuum probe is shown in the supplementary materials Figure~\ref{Afig:reflectivity}. The $\Delta R / R$ response displays an almost instantaneous decrease of the reflectivity and at least two time constants ($\sim 5$\,ps and $\sim 200$\,ps) in the recovery, as estimated by exponential fitting. The $\Delta R / R$ evolution with the delay can be due to several phenomena, such as the presence of free electrons above the Fermi level that modify the Coulomb screening mechanisms \cite{Granas_NiO}. Moreover, the positive $\Delta R / R$ signal after 10\,ps at lower energies ($\sim 3.5$\,eV) might indicate the transient heating of the atomic and spin lattices, by the transfer of energy from electrons to phonons and magnons. Therefore, the transient reflectivity results suggest that thermal effects become relevant only after 10\,ps, and that the softening of the $dd_1$ peak in the first 10\,ps in \textit{tr}RIXS is likely not due to thermalization. Its persistence at longer delays can have either thermal or electronic origin and further investigations are needed to solve the dilemma. 

By combining the dynamics of the EG peak intensity, the $dd_1$ softening and the transient reflectivity measurements we have thus the evidence that CT excitons are created by the absorption of 4.66\,eV photons; they then decay within a couple of picoseconds while delocalized charges, created either directly or from the decay of the CT excitons, persist at least up to $10$ ps; at longer timescales, thermal effects might become dominant. 
Besides the role of the CT excitons, what remains unclear is the nature of the delocalized states. 
X-ray spectroscopies clearly show that after 2\,ps the electrons photo-excited from the oxygen $2p$ band are not residing on localized $3d$ states anymore. This is supported both by our results and by the $tr$XAS measurements in ref. \cite{Eschenlohr_NiO_CT_XAS}, where it is clear that the $tr$XAS pre-edge disappears in few ps. The photo-excited electrons are likely populating the Ni $4sp$ band, being spatially decoupled from the holes in the O $2p$ band. Additionally, we cannot exclude that in-gap states of mixed $dd$ and CT excitonic character \cite{acharya2023color, Sokolov_2012} can trap the electrons released from the CT excitons and mediate the eventual relaxation to the ground state. Further insight will come from \textit{tr}RIXS by tuning the experimental parameters. By decreasing the energy of the pump photons to match the optical gap, and below, one could check for the persistence of the CT exciton when the photo-excitation process is made at threshold and not above it. By working at the oxygen K edge one could measure \textit{tr}RIXS spectra more sensitive to the CT continuum above 4\,eV and thus possibly be able to determine the actual time scale of the electron-hole recombination. By measuring spectra with higher energy resolution one could access the magnon and phonon excitation peaks to investigate their role in the thermalization process. This is especially relevant because in correlated insulators the photo-excited states above the optical gap are often magnetically dressed, for both Mott-Hubbard \cite{Huang_spinMott,mehio2023hubbard,Murakami_spinMott,dean2016Ultrafast} and CT insulators \cite{belvin2021exciton,kang2020coherent,gillmeister2020ultrafast}. 

Our findings on the excitonic nature of photo-excitation can be extended to other correlated transition-metal oxides, encompassing not only those with a wide optical gap but also doped systems, such as high-$T_c$ superconducting cuprates. Notably, the undoped parent compounds of cuprates fall within the class of CT insulators. Although starting from an analogous photo-excited state, in metallic systems such as doped cuprates the de-excitation pathways can differ significantly from those of the parent insulating compounds \cite{giannetti2016review}, resulting in distinct dynamics of the various spectral features and, consequently, altering the interplay of the different degrees of freedom involved. However, it is important to note that for cuprates the detection of CT excitons by Cu L$_3$ \textit{tr}RIXS is very difficult if not impossible. Indeed, starting from the $3d^9$ ground state the CT exciton would have $3d^{10}$\underline{L} configuration, which is invisible to XAS and RIXS at the Cu L edges because of the lack of resonance for the $2p\to 3d$ transition. Therefore, while a different phenomenology of \textit{tr}RIXS in cuprates and NiO has to be expected and a distinct approach to the interpretation of experimental results is probably needed, our results on the formation of the CT exciton in NiO can serve as reference when the CT exciton cannot be directly observed. Besides the peculiar case of cuprates, the method presented here for NiO can be applied to many other transition-metal oxides with more exotic electronic and magnetic properties and, more in general, \textit{tr}RIXS can be used as a powerful new tool to decipher the photo-induced processes in these quantum materials.


\newpage

\clearpage 

%
\bibliography{sn-bibliography}

\begin{thebibliography}{10}
\providecommand{\url}[1]{\texttt{#1}}
\expandafter\ifx\csname urlstyle\endcsname\relax
  \providecommand{\doi}[1]{doi:\discretionary{}{}{}#1}\else
  \providecommand{\doi}{doi:\discretionary{}{}{}\begingroup \urlstyle{rm}\Url}\fi

\bibitem{cavalleri2001femtosecond}
A.~Cavalleri, \emph{et~al.}, {Femtosecond Structural Dynamics in ${\mathrm{VO}}_{2}$ during an Ultrafast Solid-Solid Phase Transition}. \emph{Phys. Rev. Lett.} \textbf{87}, 237401 (2001), \doi{10.1103/PhysRevLett.87.237401}, \url{https://link.aps.org/doi/10.1103/PhysRevLett.87.237401}.

\bibitem{polli2007Coherent}
D.~Polli, \emph{et~al.}, {Coherent Orbital Waves in the Photo-Induced Insulator{\textendash}Metal Dynamics of a Magnetoresistive Manganite}. \emph{Nature Materials} \textbf{6}~(9), 643--647 (2007), \doi{10.1038/nmat1979}, \url{https://www.nature.com/articles/nmat1979}.

\bibitem{boschini2018Collapse}
F.~Boschini, \emph{et~al.}, {Collapse of Superconductivity in Cuprates via Ultrafast Quenching of Phase Coherence}. \emph{Nature Materials} \textbf{17}~(5), 416--420 (2018), \doi{10.1038/s41563-018-0045-1}, \url{https://www.nature.com/articles/s41563-018-0045-1}.

\bibitem{Nevola_meleted_Fe_based}
D.~Nevola, \emph{et~al.}, {Ultrafast Melting of Superconductivity in an Iron-Based Superconductor}. \emph{Phys. Rev. X} \textbf{13}, 011001 (2023), \doi{10.1103/PhysRevX.13.011001}, \url{https://link.aps.org/doi/10.1103/PhysRevX.13.011001}.

\bibitem{mitrano2019ultrafast}
M.~Mitrano, \emph{et~al.}, {Ultrafast time-resolved x-ray scattering reveals diffusive charge order dynamics in ${\mathrm{La}}_{2-x}{\mathrm{Ba}}_{x}\mathrm{Cu}{\mathrm{O}}_{4}$}. \emph{Science Advances} \textbf{5}~(8), eaax3346 (2019), \doi{10.1126/sciadv.aax3346}, \url{https://www.science.org/doi/abs/10.1126/sciadv.aax3346}.

\bibitem{dean2016Ultrafast}
M.~P.~M. Dean, \emph{et~al.}, {Ultrafast Energy- and Momentum-Resolved Dynamics of Magnetic Correlations in the Photo-Doped {{Mott}} Insulator ${\mathrm{Sr}}_{2}\mathrm{Ir}{\mathrm{O}}_{4}$}. \emph{Nature Materials} \textbf{15}~(6), 601--605 (2016), \doi{10.1038/nmat4641}, \url{https://www.nature.com/articles/nmat4641}.

\bibitem{Mitrano_XAS_U_renorm}
D.~R. Baykusheva, \emph{et~al.}, {Ultrafast Renormalization of the On-Site Coulomb Repulsion in a Cuprate Superconductor}. \emph{Phys. Rev. X} \textbf{12}, 011013 (2022), \doi{10.1103/PhysRevX.12.011013}, \url{https://link.aps.org/doi/10.1103/PhysRevX.12.011013}.

\bibitem{Eschenlohr_NiO_CT_XAS}
{Lojewski, Tobias and Gole\ifmmode \check{z}\else \v{z}\fi{}, Denis and Ollefs, Katharina and Le Guyader, Lo\"{\i}c and K\"ammerer, Lea and Rothenbach, Nico and Engel, Robin Y. and Miedema, Piter S. and Beye, Martin and Chiuzb\ifmmode \u{a}\else \u{a}\fi{}ian, Gheorghe S. and Carley, Robert and Gort, Rafael and Van Kuiken, Benjamin E. and Mercurio, Giuseppe and Schlappa, Justina and Yaroslavtsev, Alexander and Scherz, Andreas and D\"oring, Florian and David, Christian and Wende, Heiko and Bovensiepen, Uwe and Eckstein, Martin and Werner, Philipp and Eschenlohr, Andrea}, {Photoinduced charge transfer renormalization in NiO}. \emph{Phys. Rev. B} \textbf{110}, 245120 (2024), \doi{10.1103/PhysRevB.110.245120}, \url{https://link.aps.org/doi/10.1103/PhysRevB.110.245120}.

\bibitem{Granas_NiO}
O.~Gr\aa{}n\"as, \emph{et~al.}, Ultrafast modification of the electronic structure of a correlated insulator. \emph{Phys. Rev. Res.} \textbf{4}, L032030 (2022), \doi{10.1103/PhysRevResearch.4.L032030}, \url{https://link.aps.org/doi/10.1103/PhysRevResearch.4.L032030}.

\bibitem{ZSA}
J.~Zaanen, G.~A. Sawatzky, J.~W. Allen, Band gaps and electronic structure of transition-metal compounds. \emph{Phys. Rev. Lett.} \textbf{55}, 418--421 (1985), \doi{10.1103/PhysRevLett.55.418}, \url{https://link.aps.org/doi/10.1103/PhysRevLett.55.418}.

\bibitem{Clarke_MHExc}
{Clarke, David G.}, {Particle-hole bound states in Mott-Hubbard insulators}. \emph{{Phys. Rev. B}} \textbf{48}, 7520--7525 (1993), \doi{10.1103/PhysRevB.48.7520}, \url{https://link.aps.org/doi/10.1103/PhysRevB.48.7520}.

\bibitem{Lenarcic_MHExc}
{Lenar\ifmmode \check{c}\else \v{c}\fi{}i\ifmmode \check{c}\else \v{c}\fi{}, Zala and Prelov\ifmmode \check{s}\else \v{s}\fi{}ek, Peter}, {Ultrafast Charge Recombination in a Photoexcited Mott-Hubbard Insulator}. \emph{{Phys. Rev. Lett.}} \textbf{111}, 016401 (2013), \doi{10.1103/PhysRevLett.111.016401}, \url{https://link.aps.org/doi/10.1103/PhysRevLett.111.016401}.

\bibitem{Huang_MHE}
{Huang, T.-S. and Baldwin, C. L. and Hafezi, M. and Galitski, V.}, {Spin-mediated Mott excitons}. \emph{{Phys. Rev. B}} \textbf{107}, 075111 (2023), \doi{10.1103/PhysRevB.107.075111}, \url{https://link.aps.org/doi/10.1103/PhysRevB.107.075111}.

\bibitem{Shinjo_MHE}
{Shinjo, Kazuya and Tamaki, Yoshiki and Sota, Shigetoshi and Tohyama, Takami}, {Density-matrix renormalization group study of optical conductivity of the Mott insulator for two-dimensional clusters}. \emph{{Phys. Rev. B}} \textbf{104}, 205123 (2021), \doi{10.1103/PhysRevB.104.205123}, \url{https://link.aps.org/doi/10.1103/PhysRevB.104.205123}.

\bibitem{Wrobel_MHE}
{Wr\'obel, P. and Eder, R.}, {Excitons in Mott insulators}. \emph{{Phys. Rev. B}} \textbf{66}, 035111 (2002), \doi{10.1103/PhysRevB.66.035111}, \url{https://link.aps.org/doi/10.1103/PhysRevB.66.035111}.

\bibitem{mehio2023hubbard}
{Mehio, Omar and Li, Xinwei and Ning, Honglie and Lenar{\v{c}}i{\v{c}}, Zala and Han, Yuchen and Buchhold, Michael and Porter, Zach and Laurita, Nicholas J and Wilson, Stephen D and Hsieh, David}, {A Hubbard exciton fluid in a photo-doped antiferromagnetic Mott insulator}. \emph{Nature Physics} \textbf{19}~(12), 1876--1882 (2023), \url{https://doi.org/10.1038/s41567-023-02204-2}.

\bibitem{Hanamura_CTE}
{Hanamura, Eiichi and Dan, Nguyen Trung and Tanabe, Yukito}, {Excitons and two-magnon Raman scattering of the strongly correlated systems ${\mathrm{La}}_{2}{\mathrm{CuO}}_{4}$ and ${\mathrm{YBa}}_{2}{\mathrm{Cu}}_{3}{\mathrm{O}}_{6}$}. \emph{{Phys. Rev. B}} \textbf{62}, 7033--7044 (2000), \doi{10.1103/PhysRevB.62.7033}, \url{https://link.aps.org/doi/10.1103/PhysRevB.62.7033}.

\bibitem{Ament_review}
L.~J.~P. Ament, M.~van Veenendaal, T.~P. Devereaux, J.~P. Hill, J.~van~den Brink, Resonant inelastic x-ray scattering studies of elementary excitations. \emph{Rev. Mod. Phys.} \textbf{83}, 705--767 (2011), \doi{10.1103/RevModPhys.83.705}, \url{https://link.aps.org/doi/10.1103/RevModPhys.83.705}.

\bibitem{Mitrano_review}
{Mitrano, M. and Johnston, S. and Kim, Young-June and Dean, M. P. M.}, {Exploring Quantum Materials with Resonant Inelastic X-Ray Scattering}. \emph{Phys. Rev. X} \textbf{14}, 040501 (2024), \doi{10.1103/PhysRevX.14.040501}, \url{https://link.aps.org/doi/10.1103/PhysRevX.14.040501}.

\bibitem{Ghiringhelli_2005}
G.~Ghiringhelli, \emph{et~al.}, {NiO as a test case for high resolution resonant inelastic soft x-ray scattering}. \emph{Journal of Physics: Condensed Matter} \textbf{17}~(35), 5397 (2005), \doi{10.1088/0953-8984/17/35/007}, \url{https://dx.doi.org/10.1088/0953-8984/17/35/007}.

\bibitem{Moretti_Sala_2011}
M.~Moretti~Sala, \emph{et~al.}, {Energy and symmetry of dd excitations in undoped layered cuprates measured by Cu L3 resonant inelastic x-ray scattering}. \emph{New Journal of Physics} \textbf{13}~(4), 043026 (2011), \doi{10.1088/1367-2630/13/4/043026}, \url{https://dx.doi.org/10.1088/1367-2630/13/4/043026}.

\bibitem{Schlappa_orbitons_2012}
J.~Schlappa, \emph{et~al.}, {Spin-orbital separation in the quasi-one-dimensional Mott insulator Sr$_2$CuO$_3$}. \emph{Nature} \textbf{485}~(7396), 82--85 (2012), \url{https://doi.org/10.1038/nature10974}.

\bibitem{hepting2020Electronica}
M.~Hepting, \emph{et~al.}, {Electronic Structure of the Parent Compound of Superconducting Infinite-Layer Nickelates}. \emph{Nature Materials} \textbf{19}~(4), 381--385 (2020), \doi{10.1038/s41563-019-0585-z}, \url{https://www.nature.com/articles/s41563-019-0585-z}.

\bibitem{braicovich2010magnetic}
L.~Braicovich, \emph{et~al.}, {Magnetic Excitations and Phase Separation in the Underdoped ${\mathrm{La}}_{2\ensuremath{-}x}{\mathrm{Sr}}_{x}{\mathrm{CuO}}_{4}$ Superconductor Measured by Resonant Inelastic X-Ray Scattering}. \emph{Phys. Rev. Lett.} \textbf{104}, 077002 (2010), \doi{10.1103/PhysRevLett.104.077002}, \url{https://link.aps.org/doi/10.1103/PhysRevLett.104.077002}.

\bibitem{LeTacon_paramag_2011}
M.~Le~Tacon, \emph{et~al.}, {Intense paramagnon excitations in a large family of high-temperature superconductors}. \emph{Nat. Phys.} \textbf{7}~(9), 725--730 (2011), \url{https://doi.org/10.1038/nphys2041}.

\bibitem{PengInfluence_2017}
Y.~Peng, \emph{et~al.}, {Influence of apical oxygen on the extent of in-plane exchange interaction in cuprate superconductors}. \emph{Nat. Phys.} \textbf{13}~(12), 1201--1206 (2017), \url{https://doi.org/10.1038/nphys4248}.

\bibitem{WSL_ILN_magnons_2021}
H.~Lu, \emph{et~al.}, Magnetic excitations in infinite-layer nickelates. \emph{Science} \textbf{373}~(6551), 213--216 (2021), \doi{10.1126/science.abd7726}, \url{https://www.science.org/doi/abs/10.1126/science.abd7726}.

\bibitem{Rossi_phonons_2019}
M.~Rossi, \emph{et~al.}, {Experimental Determination of Momentum-Resolved Electron-Phonon Coupling}. \emph{Phys. Rev. Lett.} \textbf{123}, 027001 (2019), \doi{10.1103/PhysRevLett.123.027001}, \url{https://link.aps.org/doi/10.1103/PhysRevLett.123.027001}.

\bibitem{Vale_iridate_phonons_2019}
J.~G. Vale, \emph{et~al.}, High-resolution resonant inelastic x-ray scattering study of the electron-phonon coupling in honeycomb $\ensuremath{\alpha}\ensuremath{-}{\mathrm{Li}}_{2}{\mathrm{IrO}}_{3}$. \emph{Phys. Rev. B} \textbf{100}, 224303 (2019), \doi{10.1103/PhysRevB.100.224303}, \url{https://link.aps.org/doi/10.1103/PhysRevB.100.224303}.

\bibitem{chaix2017dispersive}
L.~Chaix, \emph{et~al.}, {Dispersive charge density wave excitations in Bi$_2$Sr$_2$CaCu$_2$O$_{8+\delta}$}. \emph{Nat. Phys.} \textbf{13}~(10), 952--956 (2017), \doi{https://doi.org/10.1038/nphys4157}.

\bibitem{lee2021spectroscopic}
W.-S. Lee, \emph{et~al.}, {Spectroscopic fingerprint of charge order melting driven by quantum fluctuations in a cuprate}. \emph{Nat. Phys.} \textbf{17}~(1), 53--57 (2021), \doi{https://doi.org/10.1038/s41567-020-0993-7}.

\bibitem{ArpaiaScience_2019}
R.~Arpaia, \emph{et~al.}, {Dynamical charge density fluctuations pervading the phase diagram of a Cu-based high Tc superconductor}. \emph{Science} \textbf{365}~(6456), 906--910 (2019), \url{https://www.science.org/doi/abs/10.1126/science.aav1315}.

\bibitem{HuangCDF_2021}
H.~Y. Huang, \emph{et~al.}, Quantum Fluctuations of Charge Order Induce Phonon Softening in a Superconducting Cuprate. \emph{Phys. Rev. X} \textbf{11}, 041038 (2021), \doi{10.1103/PhysRevX.11.041038}, \url{https://link.aps.org/doi/10.1103/PhysRevX.11.041038}.

\bibitem{Wrobel_Mottness_NiO}
F.~Wrobel, \emph{et~al.}, {Doped NiO: The mottness of a charge transfer insulator}. \emph{Phys. Rev. B} \textbf{101}, 195128 (2020), \doi{10.1103/PhysRevB.101.195128}, \url{https://link.aps.org/doi/10.1103/PhysRevB.101.195128}.

\bibitem{newman1959optical}
R.~Newman, R.~M. Chrenko, {Optical Properties of Nickel Oxide}. \emph{Phys. Rev.} \textbf{114}, 1507--1513 (1959), \doi{10.1103/PhysRev.114.1507}, \url{https://link.aps.org/doi/10.1103/PhysRev.114.1507}.

\bibitem{finazzi_resAES}
M.~Finazzi, N.~B. Brookes, F.~De~Groot, {2$p$3$s$3$p$, 2$p$3$p$3$p$, and 2$p$3$s$3$s$ resonant Auger spectroscopy from NiO}. \emph{Physical Review B} \textbf{59}~(15), 9933 (1999), \url{https://journals.aps.org/prb/abstract/10.1103/PhysRevB.59.9933}.

\bibitem{Tjernberg_resonant_PES_NiO}
O.~Tjernberg, \emph{et~al.}, {Resonant photoelectron spectroscopy on NiO}. \emph{Phys. Rev. B} \textbf{53}, 10372--10376 (1996), \doi{10.1103/PhysRevB.53.10372}, \url{https://link.aps.org/doi/10.1103/PhysRevB.53.10372}.

\bibitem{Sawatzky_DFT_NiO}
V.~I. Anisimov, I.~V. Solovyev, M.~A. Korotin, M.~T. Czy\ifmmode~\dot{z}\else \.{z}\fi{}yk, G.~A. Sawatzky, {Density-functional theory and NiO photoemission spectra}. \emph{Phys. Rev. B} \textbf{48}, 16929--16934 (1993), \doi{10.1103/PhysRevB.48.16929}, \url{https://link.aps.org/doi/10.1103/PhysRevB.48.16929}.

\bibitem{ghiringhelli2009observation}
G.~Ghiringhelli, \emph{et~al.}, {Observation of Two Nondispersive Magnetic Excitations in NiO by Resonant Inelastic Soft-X-Ray Scattering}. \emph{Phys. Rev. Lett.} \textbf{102}, 027401 (2009), \doi{10.1103/PhysRevLett.102.027401}, \url{https://link.aps.org/doi/10.1103/PhysRevLett.102.027401}.

\bibitem{betto2017three}
D.~Betto, \emph{et~al.}, {Three-dimensional dispersion of spin waves measured in NiO by resonant inelastic x-ray scattering}. \emph{Phys. Rev. B} \textbf{96}, 020409 (2017), \doi{10.1103/PhysRevB.96.020409}, \url{https://link.aps.org/doi/10.1103/PhysRevB.96.020409}.

\bibitem{nag2020many}
A.~Nag, \emph{et~al.}, {Many-Body Physics of Single and Double Spin-Flip Excitations in NiO}. \emph{Phys. Rev. Lett.} \textbf{124}, 067202 (2020), \doi{10.1103/PhysRevLett.124.067202}, \url{https://link.aps.org/doi/10.1103/PhysRevLett.124.067202}.

\bibitem{Hufner_NiO}
S.~Hüfner, {Electronic structure of NiO and related 3d-transition-metal compounds}. \emph{Advances in Physics} \textbf{43}~(2), 183--356 (1994), \doi{10.1080/00018739400101495}, \url{https://doi.org/10.1080/00018739400101495}.

\bibitem{Tzschaschel_PhysRevB.95.174407}
{Tzschaschel, Christian and Otani, Kensuke and Iida, Ryugo and Shimura, Tsutomu and Ueda, Hiroaki and G\"unther, Stefan and Fiebig, Manfred and Satoh, Takuya}, {Ultrafast optical excitation of coherent magnons in antiferromagnetic NiO}. \emph{Phys. Rev. B} \textbf{95}, 174407 (2017), \doi{10.1103/PhysRevB.95.174407}, \url{https://link.aps.org/doi/10.1103/PhysRevB.95.174407}.

\bibitem{Takahara_PhysRevB.86.094301}
{Takahara, M. and Jinn, H. and Wakabayashi, S. and Moriyasu, T. and Kohmoto, T.}, {Observation of coherent acoustic phonons and magnons in an antiferromagnet NiO}. \emph{Phys. Rev. B} \textbf{86}, 094301 (2012), \doi{10.1103/PhysRevB.86.094301}, \url{https://link.aps.org/doi/10.1103/PhysRevB.86.094301}.

\bibitem{Satoh_PhysRevB.74.012404}
{Satoh, Takuya and Duong, Nguyen Phuc and Fiebig, Manfred}, {Coherent control of antiferromagnetism in $\mathrm{NiO}$}. \emph{Phys. Rev. B} \textbf{74}, 012404 (2006), \doi{10.1103/PhysRevB.74.012404}, \url{https://link.aps.org/doi/10.1103/PhysRevB.74.012404}.

\bibitem{Satoh_PhysRevLett.105.077402}
{Satoh, Takuya and Cho, Sung-Jin and Iida, Ryugo and Shimura, Tsutomu and Kuroda, Kazuo and Ueda, Hiroaki and Ueda, Yutaka and Ivanov, B. A. and Nori, Franco and Fiebig, Manfred}, {Spin Oscillations in Antiferromagnetic NiO Triggered by Circularly Polarized Light}. \emph{Phys. Rev. Lett.} \textbf{105}, 077402 (2010), \doi{10.1103/PhysRevLett.105.077402}, \url{https://link.aps.org/doi/10.1103/PhysRevLett.105.077402}.

\bibitem{Duong_PhysRevLett.93.117402}
{Duong, N. P. and Satoh, T. and Fiebig, M.}, {Ultrafast Manipulation of Antiferromagnetism of NiO}. \emph{Phys. Rev. Lett.} \textbf{93}, 117402 (2004), \doi{10.1103/PhysRevLett.93.117402}, \url{https://link.aps.org/doi/10.1103/PhysRevLett.93.117402}.

\bibitem{Bossini_PhysRevLett.127.077202}
{Bossini, D. and Pancaldi, M. and Soumah, L. and Basini, M. and Mertens, F. and Cinchetti, M. and Satoh, T. and Gomonay, O. and Bonetti, S.}, {Ultrafast Amplification and Nonlinear Magnetoelastic Coupling of Coherent Magnon Modes in an Antiferromagnet}. \emph{Phys. Rev. Lett.} \textbf{127}, 077202 (2021), \doi{10.1103/PhysRevLett.127.077202}, \url{https://link.aps.org/doi/10.1103/PhysRevLett.127.077202}.

\bibitem{gillmeister2020ultrafast}
{Gillmeister, Konrad and Gole{\v{z}}, Denis and Chiang, Cheng-Tien and Bittner, Nikolaj and Pavlyukh, Yaroslav and Berakdar, Jamal and Werner, Philipp and Widdra, Wolf}, {Ultrafast coupled charge and spin dynamics in strongly correlated NiO}. \emph{Nature communications} \textbf{11}~(1), 4095 (2020), \url{https://doi.org/10.1038/s41467-020-17925-8}.

\bibitem{Kunnus_2016}
{Kunnus, Kristjan and Josefsson, Ida and Rajkovic, Ivan and Schreck, Simon and Quevedo, Wilson and Beye, Martin and Grübel, Sebastian and Scholz, Mirko and Nordlund, Dennis and Zhang, Wenkai and Hartsock, Robert W and Gaffney, Kelly J and Schlotter, William F and Turner, Joshua J and Kennedy, Brian and Hennies, Franz and Techert, Simone and Wernet, Philippe and Odelius, Michael and Föhlisch, Alexander}, {Anti-Stokes resonant x-ray Raman scattering for atom specific and excited state selective dynamics}. \emph{New Journal of Physics} \textbf{18}~(10), 103011 (2016), \doi{10.1088/1367-2630/18/10/103011}, \url{https://dx.doi.org/10.1088/1367-2630/18/10/103011}.

\bibitem{quanty_paper}
{Haverkort, M. W. and Zwierzycki, M. and Andersen, O. K.}, {Multiplet ligand-field theory using Wannier orbitals}. \emph{Phys. Rev. B} \textbf{85}, 165113 (2012), \doi{10.1103/PhysRevB.85.165113}, \url{https://link.aps.org/doi/10.1103/PhysRevB.85.165113}.

\bibitem{Huang_dd_screening}
S.-W. Huang, \emph{et~al.}, {Precise $dd$ excitations and commensurate intersite Coulomb interactions in the dissimilar cuprates $\mathrm{Y}{\mathrm{Ba}}_{2}{\mathrm{Cu}}_{3}{\mathrm{O}}_{7-y}$ and ${\mathrm{La}}_{2-x}{\mathrm{Sr}}_{x}\mathrm{Cu}{\mathrm{O}}_{4}$}. \emph{Phys. Rev. B} \textbf{107}, 134513 (2023), \doi{10.1103/PhysRevB.107.134513}, \url{https://link.aps.org/doi/10.1103/PhysRevB.107.134513}.

\bibitem{merzoni2023charge}
{Merzoni, Giacomo and Martinelli, Leonardo and Braicovich, Lucio and Brookes, Nicholas B. and Lombardi, Floriana and Rosa, Francesco and Arpaia, Riccardo and Moretti Sala, Marco and Ghiringhelli, Giacomo}, {Charge response function probed by resonant inelastic x-ray scattering: Signature of electronic gaps of ${\mathrm{YBa}}_{2}{\mathrm{Cu}}_{3}{\mathrm{O}}_{7\ensuremath{-}\ensuremath{\delta}}$}. \emph{Phys. Rev. B} \textbf{109}, 184506 (2024), \doi{10.1103/PhysRevB.109.184506}, \url{https://link.aps.org/doi/10.1103/PhysRevB.109.184506}.

\bibitem{fumagalli2019polarization}
R.~Fumagalli, \emph{et~al.}, {Polarization-resolved Cu ${L}_{3}$-edge resonant inelastic x-ray scattering of orbital and spin excitations in ${\mathrm{NdBa}}_{2}{\mathrm{Cu}}_{3}{\mathrm{O}}_{7\ensuremath{-}\ensuremath{\delta}}$}. \emph{Phys. Rev. B} \textbf{99}, 134517 (2019), \doi{10.1103/PhysRevB.99.134517}, \url{https://link.aps.org/doi/10.1103/PhysRevB.99.134517}.

\bibitem{barantaniPRXdd}
{Barantani, F. and Tran, M. K. and Madan, I. and Kapon, I. and Bachar, N. and Asmara, T. C. and Paris, E. and Tseng, Y. and Zhang, W. and Hu, Y. and Giannini, E. and Gu, G. and Devereaux, T. P. and Berthod, C. and Carbone, F. and Schmitt, T. and van der Marel, D.}, {Resonant Inelastic X-Ray Scattering Study of Electron-Exciton Coupling in High-${T}_{c}$ Cuprates}. \emph{Phys. Rev. X} \textbf{12}, 021068 (2022), \doi{10.1103/PhysRevX.12.021068}, \url{https://link.aps.org/doi/10.1103/PhysRevX.12.021068}.

\bibitem{giannetti2016review}
{Claudio Giannetti, Massimo Capone, Daniele Fausti, Michele Fabrizio, Fulvio Parmigiani and Dragan Mihailovic}, {Ultrafast optical spectroscopy of strongly correlated materials and high-temperature superconductors: a non-equilibrium approach}. \emph{Advances in Physics} \textbf{65}~(2), 58--238 (2016), \doi{10.1080/00018732.2016.1194044}, \url{https://doi.org/10.1080/00018732.2016.1194044}.

\bibitem{acharya2023color}
{Acharya, Swagata and Pashov, Dimitar and Weber, Cedric and van Schilfgaarde, Mark and Lichtenstein, Alexander I and Katsnelson, Mikhail I}, {A theory for colors of strongly correlated electronic systems}. \emph{{Nature Communications}} \textbf{14}~(1), 5565 (2023), \url{https://www.nature.com/articles/s41467-023-41314-6}.

\bibitem{Sokolov_2012}
{Sokolov, V. I. and Pustovarov, V. A. and Churmanov, V. N. and Ivanov, V. Yu. and Gruzdev, N. B. and Sokolov, P. S. and Baranov, A. N. and Moskvin, A. S.}, {Unusual x-ray excited luminescence spectra of NiO suggest self-trapping of the $d$-$d$ charge-transfer exciton}. \emph{{Phys. Rev. B}} \textbf{86}, 115128 (2012), \doi{10.1103/PhysRevB.86.115128}, \url{https://link.aps.org/doi/10.1103/PhysRevB.86.115128}.

\bibitem{Huang_spinMott}
{Huang, T.-S. and Baldwin, C. L. and Hafezi, M. and Galitski, V.}, {Spin-mediated Mott excitons}. \emph{Phys. Rev. B} \textbf{107}, 075111 (2023), \doi{10.1103/PhysRevB.107.075111}, \url{https://link.aps.org/doi/10.1103/PhysRevB.107.075111}.

\bibitem{Murakami_spinMott}
{Murakami, Yuta and Takayoshi, Shintaro and Kaneko, Tatsuya and L\"auchli, Andreas M. and Werner, Philipp}, {Spin, Charge, and $\ensuremath{\eta}$-Spin Separation in One-Dimensional Photodoped Mott Insulators}. \emph{Phys. Rev. Lett.} \textbf{130}, 106501 (2023), \doi{10.1103/PhysRevLett.130.106501}, \url{https://link.aps.org/doi/10.1103/PhysRevLett.130.106501}.

\bibitem{belvin2021exciton}
{Belvin, Carina A and Baldini, Edoardo and Ozel, Ilkem Ozge and Mao, Dan and Po, Hoi Chun and Allington, Clifford J and Son, Suhan and Kim, Beom Hyun and Kim, Jonghyeon and Hwang, Inho and others}, {Exciton-driven antiferromagnetic metal in a correlated van der Waals insulator}. \emph{Nature communications} \textbf{12}~(1), 4837 (2021), \url{https://www.nature.com/articles/s41467-021-25164-8}.

\bibitem{kang2020coherent}
{Kang, Soonmin and Kim, Kangwon and Kim, Beom Hyun and Kim, Jonghyeon and Sim, Kyung Ik and Lee, Jae-Ung and Lee, Sungmin and Park, Kisoo and Yun, Seokhwan and Kim, Taehun and others}, {Coherent many-body exciton in van der Waals antiferromagnet NiPS3}. \emph{Nature} \textbf{583}~(7818), 785--789 (2020), \url{https://doi.org/10.1038/s41586-020-2520-5}.

\bibitem{borrego_2018_applSc}
{Borrego-Varillas, Roc{\'\i}o and Ganzer, Lucia and Cerullo, Giulio and Manzoni, Cristian}, {Ultraviolet transient absorption spectrometer with sub-20-fs time resolution}. \emph{Applied Sciences} \textbf{8}~(6), 989 (2018), \url{https://www.mdpi.com/2076-3417/8/6/989}.

\bibitem{Schlappa_hRIXS}
J.~Schlappa, \emph{et~al.}, {The Heisenberg-RIXS instrument at the European XFEL}. \emph{Journal of Synchrotron Radiation} \textbf{32}~(1), 29--45 (2025), \doi{10.1107/S1600577524010890}, \url{https://doi.org/10.1107/S1600577524010890}.

\bibitem{maltezopoulos2019xgm}
T.~Maltezopoulos, \emph{et~al.}, {Operation of X-ray gas monitors at the European XFEL}. \emph{Journal of Synchrotron Radiation} \textbf{26}~(4), 1045--1051 (2019), \doi{10.1107/S1600577519003795}, \url{https://doi.org/10.1107/S1600577519003795}.

\bibitem{gerasimova_mono}
N.~Gerasimova, \emph{et~al.}, {The soft X-ray monochromator at the SASE3 beamline of the European XFEL: from design to operation}. \emph{Journal of Synchrotron Radiation} \textbf{29}~(5), 1299--1308 (2022), \doi{10.1107/S1600577522007627}, \url{https://doi.org/10.1107/S1600577522007627}.

\bibitem{carley2022scs_review_report}
{Carley, Robert and Van Kuiken, Benjamin and Le Guyader, Loic and Mercurio, Giuseppe and Scherz, Andreas}, \emph{{SCS Instrument Review Report}}, Tech. rep., European X-Ray Free-Electron Laser Facility GmbH (2022), \url{https://xfel.tind.io/record/3051?v=pdf}.

\bibitem{Pergament_PPL_xfel_2014}
{M. Pergament and M. Kellert and K. Kruse and J. Wang and G. Palmer and L. Wissmann and U. Wegner and M. J. Lederer}, {High power burst-mode optical parametric amplifier with arbitrary pulse selection}. \emph{Opt. Express} \textbf{22}~(18), 22202--22210 (2014), \doi{10.1364/OE.22.022202}, \url{https://opg.optica.org/oe/abstract.cfm?URI=oe-22-18-22202}.

\bibitem{Pergament_PPL_xfel_2016}
{M. Pergament and G. Palmer and M. Kellert and K. Kruse and J. Wang and L. Wissmann and U. Wegner and M. Emons and D. Kane and G. Priebe and S. Venkatesan and T. Jezynski and F. Pallas and M. J. Lederer}, {Versatile optical laser system for experiments at the European X-ray free-electron laser facility}. \emph{Opt. Express} \textbf{24}~(26), 29349--29359 (2016), \doi{10.1364/OE.24.029349}, \url{https://opg.optica.org/oe/abstract.cfm?URI=oe-24-26-29349}.

\end{thebibliography}

\bibliographystyle{sciencemag}

%
%
%
%
%
%


\section*{Acknowledgments}
GM, LMa, MMS and GG thank Lucio Braicovich, Nick Brookes and Tom Devereaux for innumerable insightful discussions about RIXS, \textit{tr}RIXS and these results in particular.
LMa, MMS and GG acknowledge support by the projects PRIN2017 ``Quantum-2D” ID 2017Z8TS5B and PRIN2020 ``QT-FLUO`` ID 20207ZXT4Z of the Ministry for University and Research (MUR) of Italy.
YYP and GG acknowledge the support by PIK project ‘POLARIX’ by MUR.
Work by GM was jointly supported by Politecnico di Milano and European X-ray Free Electron Laser Facility GmbH.
Work at Brookhaven by MPMD was supported by the U.S. Department of Energy (DOE), Division of Materials Science, under Contract No. DE-SC0012704.
Work at Harvard by SFRT, DRB and MM was primarily supported by the U.S. Department of Energy, Office of Basic Energy Sciences, Early
Career Award Program, under Award No. DE-SC0022883.
WSL is supported by the U.S. Department of Energy, Office of Basic Energy Sciences, Materials Sciences and Engineering Division under contract No.~DE-AC02-76SF00515.
VL work was supported by Bernal Institute, Limerick, Ireland.
AA gratefully acknowledges the financial support of the Doctoral School ED 388, Chimie-Physique et Chimie Analytique de Paris Centre.
The $tr$RIXS experiment was performed during the the user assisted commissioning of the Heisenberg RIXS spectrometer hRIXS (experiment p2769). All the members of the hRIXS consortium and the SCS instrument staff are gratefully acknowledged.
We would like to thank Mr. Ulrich Setzer from Struers GmbH for polishing the NiO crystal.

\paragraph*{Author contributions:}
JS, GG, AS and MPMD conceived and coordinated the project, with the help of TS, WSL and MM.
GM, LMa, SP, SFRT, VL, LA, RC, NG, LMe, MT, BEV, ZY, SGC, TS, SS, MF, JS and GG performed the XFEL experiment at the SCS instrument of the European XFEL; AA, DRB, MK, WSL, YYP, QZQ, MM and MPMD supported remotely the XFEL experiment.
SD, OD and GC performed the transient reflectivity measurements.
GM, LMa, SP, SFRT, VL and MT implemented the data reduction routines.
GM and LMa analyzed the data and performed the ligand field calculations using the QUANTY code package written by MWH.
GM, LMa, MMS, GC, MM, MPMD, JS, AS and GG discussed and interpreted the results with inputs from MWH.
AF, JS, GG, AS, TL, ST led and coordinated the design and construction of the hRIXS spectrometer.
GM, LMa and GG wrote the manuscript with contributions by AS, GC and MF and suggestions from all the authors.

\paragraph*{Competing interests:}
The authors declare no competing interests.
\paragraph*{Data and materials availability:}
\subsection*{Supplementary materials}
Materials and Methods\\
Figures S1 to S7\\
Tables S1\\
References \textit{(63-\arabic{enumiv})}\\


\newpage


\renewcommand{\thefigure}{S\arabic{figure}}
\renewcommand{\thetable}{S\arabic{table}}
\renewcommand{\theequation}{S\arabic{equation}}
\renewcommand{\thepage}{S\arabic{page}}
\setcounter{figure}{0}
\setcounter{table}{0}
\setcounter{equation}{0}
\setcounter{page}{1} 


\begin{center}
\section*{Supplementary Materials for\\ \scititle}

\author{
	Giacomo~Merzoni$^{1,2\ast\dagger}$,
	Leonardo~Martinelli$^{1\ast\dagger\ddagger}$,
	Sergii~Parchenko$^{2}$,\\
	Sophia~F.R.~TenHuisen$^{3}$,
        Vasily~Lebedev$^{4}$,
	Luigi~Adriano$^{2}$,
	Robert~Carley$^{2}$,\\
	Natalia~Gerasimova$^{2}$,
	Laurent~Mercadier$^{2}$,
	Martin~Teichmann$^{2}$,\\
	Benjamin~E.~van Kuiken$^{2}$,
	Zhong~Yin$^{2}$,
	Amina~Alic$^{5\S}$,
	Denitsa~R.~Baykusheva$^{3,6}$,\\
	Sorin~G.~Chiuzbaian$^{5}$,
	Stefano~Dal Conte$^{1}$,
	Oleg~Dogadov$^{1}$,
	Alexander~F\"ohlisch$^{7,8}$,\\
	Maurits~W.~Haverkort$^{9}$,
	Maximilian~Kusch$^{7}$,
	Tim~Laarmann$^{10,11}$,
	Wei-Sheng~Lee$^{12}$,\\
	Marco~Moretti Sala$^{1}$,
	Yingying~Peng$^{13}$,
	Qingzheng~Qiu$^{13}$,\\
	Thorsten~Schmitt$^{14}$,
	Sreeju~Sreekantan Nair Lalithambika$^{10}$,
	Simone~Techert$^{10,15}$,\\
	Giulio~Cerullo$^{1}$,
	Michael~F\"orst$^{16}$,
	Matteo~Mitrano$^{3}$,
	Mark~P.M.~Dean$^{17}$,\\
	Justine~Schlappa$^{2}$,
	Andreas~Scherz$^{2\ast}$,
	Giacomo~Ghiringhelli$^{1,18\ast}$\\
	\small$^{1}$Dipartimento di Fisica, Politecnico di Milano, piazza Leonardo da Vinci 32, I-20133 Milano, Italy.\\
	\small$^{2}$European XFEL, Holzkoppel 4, D22869 Schenefeld, Germany.\\
	\small$^{3}$Department of Physics, Harvard University, Cambridge, Massachusetts 02138, USA.\\
        \small$^{4}$Bernal Institute, University of Limerick, Limerick, V94 T9PX, Ireland.\\
	\small$^{5}$Sorbonne Université, CNRS, Laboratoire de Chimie Physique - Matière et Rayonnement, LCPMR,\\ \small 75005 Paris, France.\\
	\small$^{6}$Institute of Science and Technology Austria, Klosterneuburg, Austria.\\
	\small$^{7}$Institute Methods and Instrumentation for Synchrotron Radiation Research, Helmholtz Center Berlin\\ \small for Materials and Energy, 12489 Berlin, Germany.\\
	\small$^{8}$Institute of Physics and Astronomy, University of Potsdam, 14476 Potsdam, Germany.\\
	\small$^{9}$Institute for Theoretical Physics, Heidelberg University, Philosophenweg 19, 69120 Heidelberg, Germany.\\
	\small$^{10}$Deutsches Elektronen-Synchrotron DESY, 22607 Hamburg, Germany.\\
	\small$^{11}$The Hamburg Centre for Ultrafast Imaging CUI, 22607 Hamburg, Germany.\\
	\small$^{12}$Stanford Institute for Materials and Energy Sciences, SLAC National Accelerator Laboratory,\\ \small2575 Sand Hill Road, Menlo Park, California 94025, USA.\\
	\small$^{13}$International Center for Quantum Materials, School of Physics, Peking University, Beijing 100871,\\ \small People's Republic of China.\\
	\small$^{14}$PSI Center for Photon Science, Paul Scherrer Institut, 5232 Villigen PSI, Switzerland.\\
	\small$^{15}$Institut für Röntgenphysik, Georg-August-Universität Göttingen, 37077 Göttingen, Germany.\\
	\small$^{16}$Max Planck Institute for the Structure and Dynamics of Matter, 22761 Hamburg, Germany.\\
	\small$^{17}$Condensed Matter Physics and Materials Science Division, Brookhaven National Laboratory, Upton,\\ \small New York 11973, USA.\\
	\small$^{18}$CNR-SPIN, Dipartimento di Fisica, Politecnico di Milano, piazza Leonardo da Vinci,\\ \small I-20133 Milano, Italy.\\
    \small$^\ast$Corresponding authors. Emails: giacomo.merzoni@polimi.it, leonardo.martinelli@physik.uzh.ch,\\ \small andreas.scherz@xfel.eu, giacomo.ghiringhelli@polimi.it\\
	\small$^\dagger$These authors contributed equally to this work.\\
    \small$^\ddagger$ Current address: Physik-Institut, Universität Zürich, Winterthurerstrasse 190, CH-8057 Zürich, Switzerland.\\
    \small$^\S$ Current address: Université Sorbonne Paris Nord and Université Paris Cité,\\ \small INSERM, LVTS, F-75018 Paris, France.
}

\end{center}


\subsubsection*{This PDF file includes:}
Materials and Methods\\
Figures S1 to S7\\
Tables S1\\

\subsection*{Materials and Methods}

\subsubsection{Sample}
The NiO sample is a 1 mm thick commercially available 100 oriented single crystal from MaTeck Material Technologie \& Kristalle GmbH. It was ground and polished using Tegramin-30 from Struers. Grinding was done in several steps using SiC paper and SiC foil of different grit size and water, in the last step \#220 FEPA P (SIC Foil \#220). Thereafter it was polished in two steps, using woven acetate cloth (MD-Dac) with water-based 3 $\mu$m diamond suspension (DiaPro Dac 3 $\mu$m) and porous neoprene cloth (MD-Chem) with 0.25 $\mu$m fumed silica suspension (OP-S NonDry).

\subsubsection{UV-visible transient reflectivity measurements}
\begin{figure}
\centering
\includegraphics[width=0.7\textwidth]{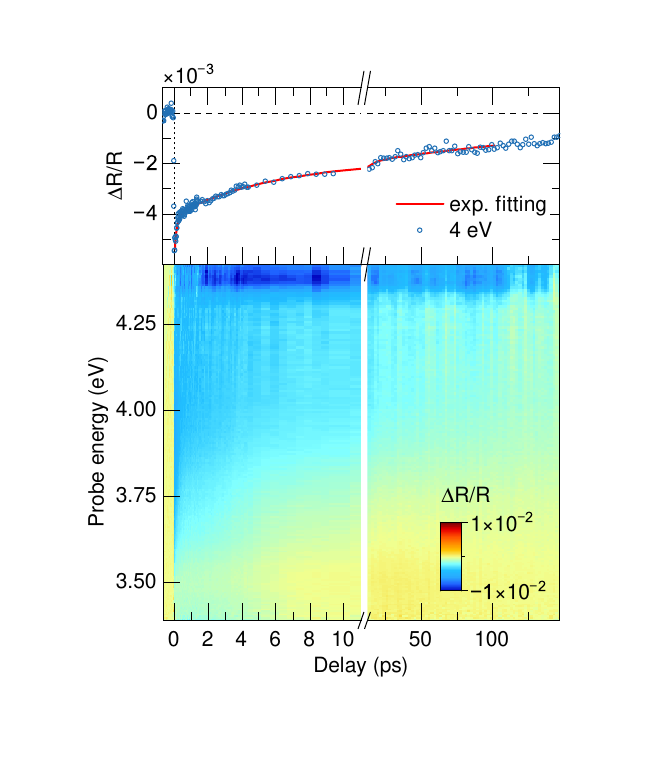}
\caption{\label{Afig:reflectivity} Transient vis-UV optical reflectivity of NiO. The 266\,nm laser pump photon energy (4.66 eV) exceeds the NiO optical gap. The pump fluence is $\sim175$ $\mu$J$/$cm$^2$. The top panel shows an example of exponential fitting, for a cut at 4 eV probe photon energy of the photon-energy/delay map of the bottom panel. Following the initial fast (sub-ps) response, the recovery does not follow a single exponential decay and time constants ranging from few ps and to hundreds of ps are apparent.}
\end{figure}

Ultrafast broadband transient reflectivity measurements were performed with a table top setup. The probe is a supercontinuum light pulse obtained by focusing the \mbox{3.1 eV} second harmonic of an amplified femtosecond Ti:sapphire laser on a 2 mm thick CaF$_2$ plate. The probe spectrum covers a wide energy region (3.4-4.4 eV) across the absorption edge of NiO. The 4.66 eV UV pump-pulse is obtained by frequency doubling a broadband few-optical-cycle visible pulse, generated by a non-collinear optical parametric amplifier (NOPA), in a thin $\beta$ barium borate (BBO) crystal, as described in detail in ref. \cite{borrego_2018_applSc}. The UV pulse is compressed down to 20 fs by a pair of prisms and temporally characterized by the two-dimensional spectral-shearing interferometry method. The pump fluence was set to  $\simeq175$ $\mu$J/cm$^2$.
A full probe photon energy-delay map of the differential reflectivity is shown in Figure \ref{Afig:reflectivity}.

\subsubsection{\textit{tr}XAS and \textit{tr}RIXS measurements}
\begin{figure}
\centering
\includegraphics[width=0.8\textwidth]{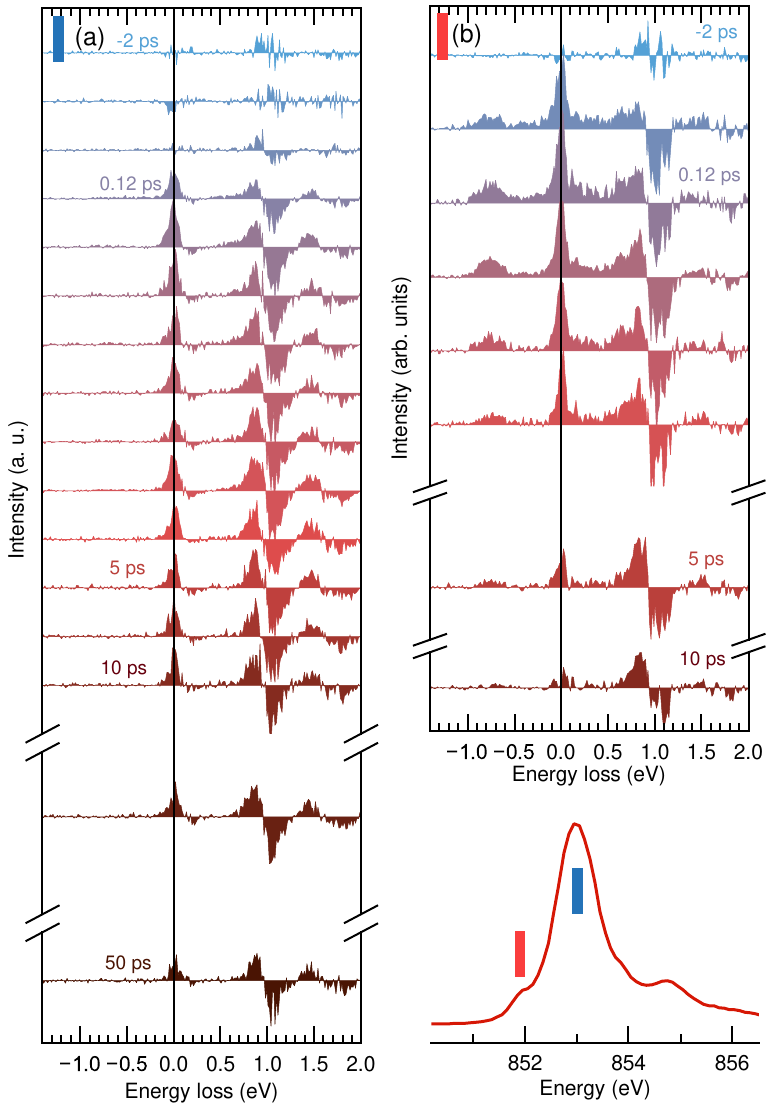}
\caption{\label{Afig:stacK_diff} Stack of the differences positive minus negative delay, for the full dataset, respectively for the main edge (panel (a)) and for the pre-edge (panel (b)). 
}
\end{figure}

\begin{figure}
\centering
\includegraphics[width=1\textwidth]{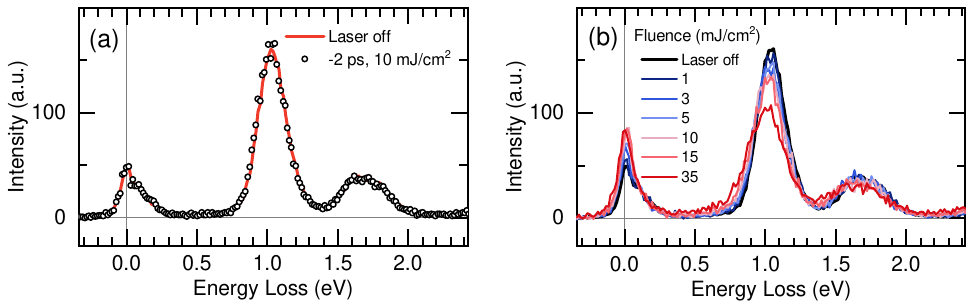}
\caption{\label{Afig:ref_and_fluence} (a) Comparison of the static (Laser off) RIXS spectrum with the one taken at -2 ps, with laser on. The almost perfect superposition demonstrates the absence of permanent heating of the sample induced by the optical laser. (b) Fluence dependence of \textit{tr}RIXS at the main edge. The delay was set to 0.5 ps. The transient modifications of the spectrum stay in a linear range at least up to 15 mJ/cm$^2$. 
}
\end{figure}

The \textit{tr}XAS and \textit{tr}RIXS experiment was performed at the SCS instrument of the European XFEL \cite{Schlappa_hRIXS}. The FEL intra train repetition rate was set to 113 kHz, i.e. 39 pulses per train, in order to match the optical laser working point. The FEL beam was attenuated to $30\%$ to avoid sample damage.
The \textit{tr}XAS data were taken in total fluorescence yield mode with a micro-channel plate (MCP) placed at 2$\theta =$ 90° and $\sim10$° above the scattering plane.
As for the RIXS spectra, a combined (beamline monochromator and hRIXS spectrometer) energy resolution of $\sim 90$ meV was achieved with a line focus of $\sim10$ $\mu$m $\times$ $\sim300$ $\mu$m. The 2$\theta$ scattering angle was kept fixed at 125° and the incidence angle was set to 66.4°, in order to have the transferred momentum \textbf{q} along the (00L) direction. The RIXS spectra are normalized by the incident x-ray intensity I$_0$, as measured by an x-rays gas monitor \cite{maltezopoulos2019xgm}. The time resolution is limited by the pulse stretching from the high-resolution monochromator grating at the Ni L$_3$ edge \cite{gerasimova_mono} and by the FEL longitudinal jitter \cite{carley2022scs_review_report}. This results in an overall time resolution of $\sim 100$ fs.
For the  optical pump-pulse, a 3.1 eV beam was obtained by frequency doubling a 1.55 eV beam, delivered as an output of a three-stage NOPA, as detailed in ref. \cite{Pergament_PPL_xfel_2014,Pergament_PPL_xfel_2016,carley2022scs_review_report}, in a BBO crystal. The 4.66 eV beam is then obtained by mixing the fundamental and the  3.1 eV beam in a second BBO crystal. The pulse duration of the fundamental was $\sim50$ fs. After the propagation of the third harmonic through an incoupling window, the resulting pulse duration can be estimated to be $\sim100$ fs. The optical laser beam is round focused at the sample position with a lens in air and it was measured by knife edge scans to be $\sim230$ $\mu$m fwhm. The calculated delivered fluence at the sample position was $\sim10$ mJ/cm$^2$. Since the measurements were performed away from normal incidence, the actual fluence on the sample was $\sim9.1$ mJ/cm$^2$. A optical laser fluence dependence in the $tr$RIXS signal is shown in Figure \ref{Afig:ref_and_fluence}(b). The relative delay between the FEL and the optical laser pulses was controlled by a linear delay stage in the fundamental optical beam path.
The spatial overlap was performed by looking at the optical and FEL beams on a pyrolitic boron nitride screen placed close to the sample. The temporal overlap was achieved first by looking at the FEL and optical laser pulses on an antenna and then refined with x-ray pump optical probe reflectivity on a SiN sample. Since the transient x-ray absorption signal from the pre-edge peak is very clear, we could check regularly the temporal overlap at the sample with the \textit{tr}XAS signal. Both spatial and temporal overlaps were checked every two hours.
The \textit{tr}RIXS acquisition was done by changing the relative delay between the FEL and the optical laser for every frame of the CCD camera of the RIXS spectrometer and then sorted afterwards. The total acquisition time for each spectrum was $\sim30$ minutes. The polarizations of the optical laser and the FEL were set to be horizontal, i.e. parallel to the scattering plane.
All the measurements were performed at room temperature.
The excitation density was estimated considering an absorption coefficient for NiO at 4.66\,eV of $\sim5\times10^5$\,cm$^{-1}$, taken from ref. \cite{newman1959optical}. The attenuation length is therefore $\sim20$\,nm. By considering that the NiO unit cell contains 4 Ni atoms and has a volume of $\simeq0.073$\,nm$^3$ and by accounting for the actual fluence of $\sim9.1$ mJ/cm$^2$ we estimate that the excited Ni sites are of the order of 10\% of the total. An overview of the full dataset for the two absorption edges is shown in Figure \ref{Afig:stacK_diff}.
In Figure \ref{Afig:ref_and_fluence}(a) a comparison between static RIXS and $tr$RIXS taken at negative delay with the nominal fluence of $\sim9.1$ mJ/cm$^2$ is shown. The absence of spectral shape or intensity modification suggest that the sample fully recovers from one optical excitation to the next.

\subsubsection{Ligand-field calculations}
As detailed in the text, the goal of the calculations is to describe the pre-edge feature of the \textit{tr}XAS and in turn the EG peak in the \textit{tr}RIXS at the pre-edge. First we optimized the parameters to make the calculations to reasonably fit the static spectrum. In order to investigate the excited states, we calculated the first 100 eigenstates of NiO, which include both the 3$d^8$ and the 3$d^9$\underline{L} configurations. As for the optically excited sites, we considered the 56$^{th}$ eigenstate, which correspond to the first excited state in the 3$d^9$\underline{L} multiplet. Details of the parameters used in the calculations (Table \ref{tableQuanty}), a comparison between static RIXS and the calculated spectrum (Figure \ref{Afig:calc_vs_exp}) and full RIXS maps for a 3$d^9$\underline{L} $^3$T$_{2g}$ and for a 3$d^8$ $^3$A$_{2g}$ initials states (Figure \ref{Afig:RIXS_map_calc}) are shown. Figure \ref{Afig:opt_calc} shows a comparison between the $tr$RIXS taken at the P edge and the calculated spectra from the 3$d^9$\underline{L} $^3$T$_{2g}$ initial state in the case of no renormalization of any parameter, with the renormalization of the charge transfer energy $\Delta_{CT}$ only and with both the renormalization of $\Delta_{CT}$ and of the crystal field parameter $10Dq$.







\begin{figure}
\centering
\includegraphics[width=0.7\textwidth]{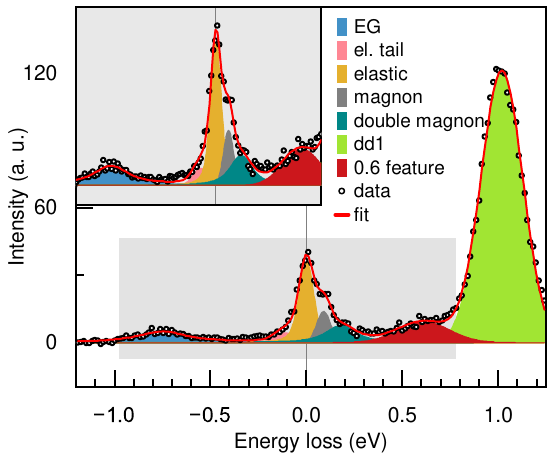}
\caption{\label{Afig:fitting} Example of fitting used to track the dynamics of the $dd_1$ softening. Gaussian profiles are used for the EG peak, the elastic, the 0.6 eV feature and the $dd_1$ peaks. Magnon and double magnon peaks are fitted with a Lorentzian profile.
}
\end{figure}

\begin{figure}
\centering
\includegraphics[width=0.7\textwidth]{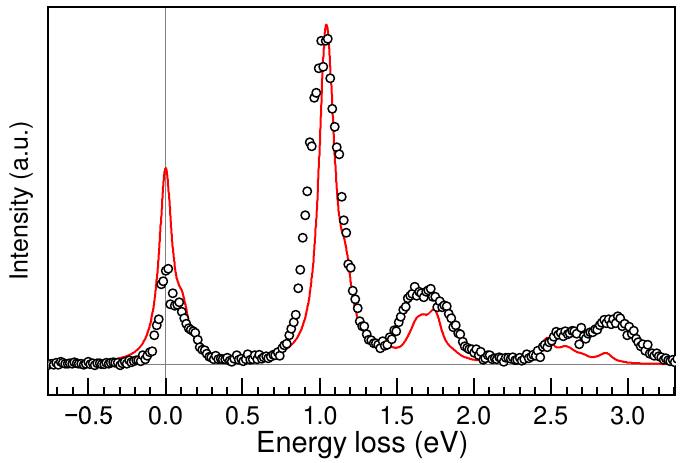}
\caption{\label{Afig:calc_vs_exp} Comparison of the ligand-field calculation of the NiO ground state, to the static RIXS at the main edge. 
}
\end{figure}

\begin{figure}
\centering
\includegraphics[width=1\textwidth]{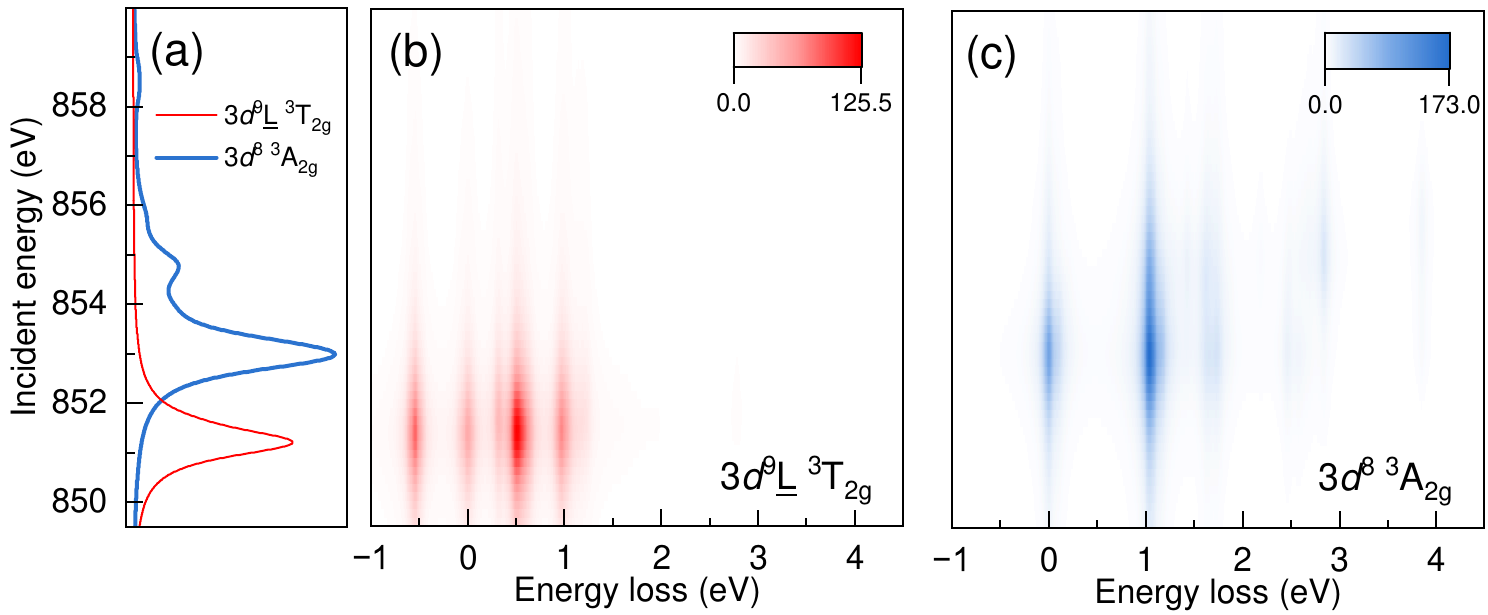}
\caption{\label{Afig:RIXS_map_calc} Overview of the ligand-field calculation. (a) calculated XAS for the 3$d^8$ $^3$A$_{2g}$ and for the 3$d^9$\underline{L} $^3$T$_{2g}$ initial states. (b,c) Full calculated RIXS maps respectively for the 3$d^9$\underline{L} $^3$T$_{2g}$ and for the 3$d^8$ $^3$A$_{2g}$ initial states.
}
\end{figure}

\begin{figure}
\centering
\includegraphics[width=0.7\textwidth]{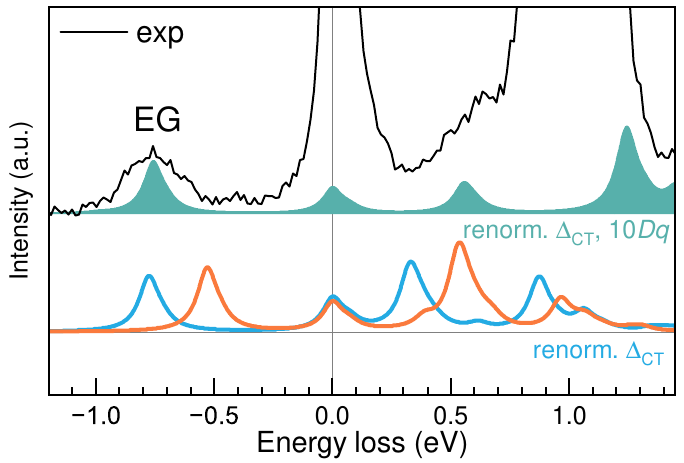}
\caption{\label{Afig:opt_calc} Optimization of the ligand-field calculation parameters to quantitatively match the energy of the two rising peaks in the experimental \textit{tr}RIXS at the pre-edge, at 0.3\,ps delay. In orange the calculated spectrum from the 3$d^9$\underline{L} $^3$T$_{2g}$ initial state is shown, as presented in Fig.\ref{fig:calculations}(c), without any renormalization of the parameters. As highlighted in the figure, the blue spectrum is calculated by re-normalizing $\Delta_{CT}$ from 4.7\,eV to 3\,eV. It can be easily see that although the EG peak energy is well matched, the 0.6 eV peak is not. The green shaded spectrum is calculated by re-normalizing both $\Delta_{CT}$ and 10$Dq$ to 3\,eV and 1.3\,eV. Although the agreement of the green spectrum is improved, such a major change of $\Delta_{CT}$ and 10$Dq$ cannot be easily explained and requires more experimental evidences to be substantiated.}
\end{figure}


\begin{sidewaystable}
\caption{Relevant parameters for the ligand-field calculations. Refer to \cite{quanty_paper} for details.}\label{Atab1}
\begin{tabular*}{\textheight}{@{\extracolsep\fill}ccccccccccccc}
\toprule%
$U_{3d,3d}$ & $U_{2p,3d}$ & $\Delta$ & $V_{eg}$ & $V_{t2g}$ & $10Dq$ & $10DqL$ & $\zeta_{3d}$ & $F^{(2)}_{dd}$ & $F^{(4)}_{dd}$ & $\zeta_{2p}$ & $F^{(2)}_{2p3d}$ \\
\midrule
7.3 & 8.5 & 4.7 & 1.8 & 1.21 & 0.75 & 1.44 & 0.081 & 7.464 & 6.87 & 11.51 & 6.67 \\
\bottomrule
\end{tabular*}
\footnotetext{The parameters have been tuned to best fit the static XAS and RIXS. Small changes have been made from the parameters listed in ref.~\cite{quanty_paper}.}
\label{tableQuanty}
\end{sidewaystable}




\end{document}